\documentclass[usegraphicx,usenatbib]{mn2e}
\usepackage{amsmath}
\usepackage{amssymb}
\usepackage{color}
\usepackage{url}

\usepackage{threeparttable}

\newcommand\lsim{\mathrel{\rlap{\lower4pt\hbox{\hskip1pt$\sim$}}
        \raise1pt\hbox{$<$}}}
\newcommand\gsim{\mathrel{\rlap{\lower4pt\hbox{\hskip1pt$\sim$}}
        \raise1pt\hbox{$>$}}}
\newcommand{\lya}{\ifmmode\mathrm{Ly}\alpha\else{}Ly$\alpha$\fi}
\newcommand{\lyb}{\ifmmode\mathrm{Ly}\beta\else{}Ly$\beta$\fi}
\newcommand{\igm}{\ifmmode\mathrm{IGM}\else{}IGM\fi}
\newcommand{\lae}{\ifmmode\mathrm{LAE}\else{}LAE\fi}
\newcommand{\h}{\ifmmode\mathrm{H}\else{}H\fi}
\newcommand{\hi}{\ifmmode\mathrm{H\,{\scriptscriptstyle I}}\else{}H\,{\scriptsize I}\fi}
\newcommand{\hii}{\ifmmode\mathrm{H\,{\scriptscriptstyle II}}\else{}H\,{\scriptsize II}\fi}
\newcommand{\heii}{\ifmmode\mathrm{He\,{\scriptscriptstyle II}}\else{}He\,{\scriptsize II}\fi}
\newcommand{\cmb}{\ifmmode\mathrm{CMB}\else{}CMB\fi}
\newcommand{\qso}{\ifmmode\mathrm{QSO}\else{}QSO\fi}
\newcommand{\eor}{\ifmmode\mathrm{EoR}\else{}EoR\fi}
\newcommand{\cmmc}{\textsc{\small 21CMMC}}
\newcommand{\cmfst}{\textsc{\small 21CMFAST}}
\newcommand{\sense}{{\small 21}\textsc{cmsense}}
\newcommand{\CH}{\textsc{cosmohammer}}
\newcommand{\emcee}{\textsc{emcee}}

\pdfoutput=1
\title[21CMMC in the epoch of heating]{Simultaneously constraining the astrophysics of reionisation and the epoch of heating with 21CMMC}

\author[B. Greig \& A. Mesinger] {Bradley~Greig$^{1}$\thanks{E-mail:~bradley.greig@sns.it} \& Andrei~Mesinger$^{1}$ \\
$^1$Scuola Normale Superiore, Piazza dei Cavalieri 7, I-56126 Pisa, Italy
}

\begin{document}
\maketitle \begin{abstract}
\noindent
The cosmic 21 cm signal is set to revolutionise our understanding of the early Universe, allowing us to probe the 3D temperature and ionisation structure of the intergalactic medium (IGM).  It will open a window onto the unseen first galaxies, showing us how their UV and X-ray photons drove the cosmic milestones of the epoch of reionisation (EoR) and epoch of heating (EoH). To facilitate parameter inference from the 21 cm signal, we previously developed \cmmc{}: a Monte Carlo Markov Chain sampler of 3D EoR simulations. Here we extend \cmmc{} to include simultaneous modelling of the EoH, resulting in a complete Bayesian inference framework for the astrophysics dominating the observable epochs of the cosmic 21 cm signal. We demonstrate that second generation interferometers, the Hydrogen Epoch of Reionisation Array (HERA) and Square Kilometre Array (SKA) will be able to constrain ionising and X-ray source properties of the first galaxies with a fractional precision of order $\sim1$--10 per cent (1$\sigma$).  The ionisation history of the Universe can be constrained to within a few percent. Using our extended framework, we quantify the bias in EoR parameter recovery incurred by the common simplification of a saturated spin temperature in the IGM. Depending on the extent of overlap between the EoR and EoH, the recovered astrophysical parameters can be biased by $\sim3-10\sigma$.
\end{abstract} 

\section{Introduction}
The 21 cm spin-flip transition of neutral hydrogen encodes a treasure trove of cosmological and astrophysical information \citep[see e.g.][]{Gnedin:1997p4494,Madau:1997p4479,Shaver:1999p4549,Tozzi:2000p4510,Gnedin:2004p4481,Furlanetto:2006p209,Morales:2010p1274,Pritchard:2012p2958}. The signal is expressed as the offset of the 21 cm brightness temperature, $\delta T_{\rm b}(\nu)$, relative to the cosmic microwave background (CMB) temperature, $T_{\rm CMB}$ \citep[e.g.][]{Furlanetto:2006p209}:
\begin{eqnarray} \label{eq:21cmTb}
\delta T_{\rm b}(\nu) &\approx& 27x_{\hi{}}(1+\delta_{\rm nl})\left(\frac{H}{{\rm d}v_{\rm r}/{\rm d}r+H}\right)
\left(1 - \frac{T_{\rm CMB}}{T_{\rm S}}\right) \nonumber \\
& & \times \left(\frac{1+z}{10}\frac{0.15}{\Omega_{\rm m}h^{2}}\right)^{1/2}
\left(\frac{\Omega_{\rm b}h^{2}}{0.023}\right)~{\rm mK},
\end{eqnarray}
where $x_{\hi{}}$ is the neutral fraction, $T_{\rm S}$ is the gas spin temperature, $\delta_{\rm nl} \equiv \rho/\bar{\rho} - 1$ is the gas overdensity, $H(z)$ is the Hubble parameter, ${\rm d}v_{\rm r}/{\rm d}r$ is the gradient of the line-of-sight component of the velocity and all quantities are evaluated at redshift $z = \nu_{0}/\nu - 1$, where $\nu_{0}$ is the 21 cm frequency.

Because equation~(\ref{eq:21cmTb}) depends on the ionisation and temperature as a function of space and time, the 21 cm signal can provide insight into the formation, growth and evolution of structure in the Universe, the nature of the first stars and galaxies and their impact on the physics of the intergalactic medium (IGM; e.g. \citealt{Barkana:2007p2929, Loeb:2013p2936, Zaroubi:2013p2976}).  The most widely-studied of these properties is the ionisation state: in the first billion years, the Universe transitioned from being nearly fully neutral to being nearly fully ionised.  This epoch of reionisation (EoR) was driven by the percolation of \hii{} regions generated by the ionising photons escaping from the first galaxies. Sourcing the $x_{\hi{}}$ term in equation~(\ref{eq:21cmTb}), the EoR should be evidenced by a rise and fall of large-scale fluctuations in the 21 cm brightness temperature (e.g. \citealt{Lidz:2008p1744}).

The second astrophysical term in equation~(\ref{eq:21cmTb}) is the IGM spin temperature,  $T_{\rm S}$. The spin temperature is thought to be regulated by the first sources of X-rays, which can heat the IGM from its post thermal decoupling adiabat to temperatures far above the CMB temperature.  While the IGM is still adiabatically cooling, the $(1 - T_{\rm CMB}/T_{\rm S})$ term in equation~(\ref{eq:21cmTb}) can be of order $-200$.  This large dynamic range means that the spatial fluctuations in temperature during this epoch of heating (EoH) can provide the strongest 21 cm signal, more than an order of magnitude larger than that during the EoR \cite[e.g.][]{Mesinger:2007p122,Pritchard:2007p3787,Baek:2010p6357,Santos:2010p6501,McQuinn:2012p3776,Mesinger:2013p1835}.  The 21 cm signal can therefore be a powerful probe of high-energy processes in the first galaxies.  The most likely of these X-ray sources are high mass X-ray binaries (HMXBs; \citealt{Power:2009p6560,Mirabel:2011p6518,Fragos:2013p6528,Power:2013p6533}) and/or the hot interstellar medium (ISM) within the first galaxies \citep[e.g.][]{Oh:2001p6617,Pacucci:2014p4323}. However, other alternative scenarios have been put forth including metal-free (Pop-III) stars \citep{Xu:2014p6671}, mini-QSOs \citep[e.g][]{Madau:2004p6564,Yue:2013p6705,Ghara:2016p6706}, dark matter annihilation \citep[e.g.][]{Cirelli:2009p6569,Evoli:2014p6707,LopezHonorez:2016p6709} or cosmic rays \citep{Leite:2017p09337}.

Numerous 21 cm experiments are currently underway, attempting to detect the cosmic 21 cm signal.  These fall into two broad categories.  The first are large-scale interferometers, seeking to detect spatial 21 cm fluctuations.  These include the Low Frequency Array (LOFAR; \citealt{vanHaarlem:2013p200,Yatawatta:2013p2980}), the Murchison Wide Field Array (MWA; \citealt{Tingay:2013p2997}), the Precision Array for Probing the Epoch of Reionisation (PAPER; \citealt{Parsons:2010p3000}), the Square Kilometre Array (SKA; \citealt{Mellema:2013p2975}) and the Hydrogen Epoch of Reionisation Array (HERA; \citealt{DeBoer:2017p6740}). The second class are single dipole or small compact array global-sky experiments seeking the volume averaged global 21 cm signal.  These include the Experiment to Detect the Global EoR Signature (EDGES; \citealt{Bowman:2010p6724}), the Sonda Cosmol\'{o}gica de las Islas para la Detecci\'{o}n de Hidr\'{o}geno Neutro (SCI-HI; \citealt{Voytek:2014p6741}), the Shaped Antenna Measurement of the Background Radio Spectrum (SARAS; \citealt{Patra:2015p6814}), Broadband Instrument for Global HydrOgen ReioNisation Signal (BIGHORNS; \citealt{Sokolowski:2015p6827}), the Large Aperture Experiment to detect the Dark Ages (LEDA; \citealt{Greenhill:2012p6829,Bernardi:2016p6834}), and the Dark Ages Radio Explorer (DARE; \citealt{Burns:2012p6941}). 

As a first step in preparation for the wealth of data expected from these 21 cm experiments, we developed a publicly available Monte Carlo Markov Chain (MCMC) EoR analysis tool \cmmc{}\footnote{https://github.com/BradGreig/21CMMC} \citep{Greig:2015p3675}. This is the first EoR analysis tool to sample 3D reionisation simulations (using \cmfst{}; \citealt{Mesinger:2007p122,Mesinger:2011p1123}) within a fully Bayesian framework for astrophysical parameter space exploration and 21 cm experiment forecasting.  In this introductory work,  we adopted the common simplifying assumption of a saturated spin temperature: $T_{\rm S} \gg T_{\rm CMB}$.  However, the applicability of this saturated limit is dependent on the poorly-known strength and spectral shape of the X-ray background in the early Universe. If the spin temperature is not fully saturated during the EoR, this could result in sizeable biases in the inferred EoR parameters \citep[e.g.][]{Watkinson2015a}. 

Beyond its impact on EoR parameter recovery, the IGM spin temperature also encodes a wealth of information on the high-energy processes in the early Universe, as mentioned above. Second generation interferometers, HERA and SKA, have the bandwidth and sensitivity to easily probe temperature fluctuations during the EoH.

In this work, we extend \cmmc{} to include a full treatment of the EoH, retaining our ability to perform on-the-fly sampling of 3D reionisation simulations. This updated \cmmc{} is capable of jointly exploring the astrophysics of both the EoR and EoH, allowing us to maximise the scientific return of upcoming second-generation telescopes.

We note that recent studies suggest that machine learning can be a viable alternative to on-the-fly MCMC sampling of \cmfst{}. \citet{Shimabukuro:2017p7140} used an artificial neural network to predict astrophysical parameters, with an accuracy of $\sim$ tens of percent.  This approach is fast, though producing Bayesian confidence limits becomes less straightforward.  Alternately, \citet{Kern:2017p8205} bypassed the on-the-fly sampling of 3D simulations by using an emulator trained on the 21 cm power spectrum. An emulator can be used in an MCMC framework, and is orders of magnitude faster at parameter forecasting compared to a direct sampling of 3D simulations (such as \cmmc{}).  This comes at the cost of $\lsim$ ten per cent in power spectrum accuracy over most of the parameter space, when the learning is performed on $\sim10^4$ \cmfst{} training samples (higher precision can be obtained by increasing the size of the training set).  Future work will test emulator accuracy on high-order summary statistics.

The remainder of this paper is organised as follows. In Section~\ref{sec:21CMMC} we summarise \cmmc{} and the associated \cmfst{} simulations used to generate 3D realisations of the cosmic 21 cm signal, outlining the updated astrophysical parameterisation to model the EoH. In Section~\ref{sec:MockObs} we introduce our mock observations to be used in our 21 cm experiment forecasting, and present the forecasts and associated discussions in Section~\ref{sec:Forecasts}. We then explore the impact of assuming the saturated IGM spin temperature limit in Section~\ref{sec:Bias}, before summarising the improvements to \cmmc{} and finishing with our closing remarks in Section~\ref{sec:conclusion}. Unless stated otherwise, we quote all quantities in comoving units and adopt the cosmological parameters: ($\Omega_\Lambda$, $\Omega_{\rm M}$, $\Omega_b$, $n$, $\sigma_8$, $H_0$) = (0.69, 0.31, 0.048, 0.97, 0.81, 68 km s$^{-1}$ Mpc$^{-1}$), consistent with recent results from the Planck mission \citep{PlanckCollaboration:2016p7780}.

\section{\cmmc{}} \label{sec:21CMMC}

In this section, we provide a short summary of the main aspects of \cmmc{}, before delving into the modelling of the cosmic 21 cm signal during the EoR and EoH in Section~\ref{sec:modelling} and developing an intuition about the full parameter set in Section~\ref{sec:description}.

\cmmc{} is a massively parallel MCMC sampler for exploring the astrophysical parameter space of reionisation. It was built using a modified version of the easy to use python module \CH{} \citep{Akeret:2012p842}, which employs the \emcee{} python module developed by \citet{ForemanMackey:2013p823} based on the affine invariant ensemble sampler of \citet{Goodman:2010p843}. At each proposed step in the computation chain, \cmmc{} performs a new, independent 3D reionisation of the 21 cm signal, using an optimised version of the publicly available \cmfst{} simulation code \citep{Mesinger:2007p122,Mesinger:2011p1123} for computational efficiency. Using a likelihood statistic (fiducially the PS), it compares the model against a mock observation generated from a larger simulation with a different set of initial conditions. For further details, we refer the reader to \citep{Greig:2015p3675}.

\subsection{Modelling the cosmic 21 cm signal with \cmfst{}} \label{sec:modelling}

In this work, we use an optimised version of the publicly-available version of \cmfst{}\textunderscore v1.1\footnote{https://github.com/andreimesinger/21cmFAST}. \cmfst{} employs approximate but efficient modelling of the underlying astrophysics of the reionisation and heating epochs.  The resulting 21 cm PS during the EoR have been found to match those of computationally expensive radiative transfer simulations to with tens of per cent on  the scales relevant to 21 cm interferometry, $\gsim 1$~Mpc \citep{Zahn:2011p1171}. We refer the reader to \citet{Mesinger:2007p122,Mesinger:2011p1123} for explicit details on the semi-numerical approach, and only provide a summary below.

\cmfst{} produces a full, 3D realisation of the 21 cm brightness temperature field, $\delta T_{\rm b}$ (c.f equation~\ref{eq:21cmTb}) which is dependent on the ionisation, density, velocity and IGM spin temperature fields. The evolved IGM density field at any redshift is obtained from an initial high resolution linear density field which is perturbed using the Zel'dovich approximation \citep{Zeldovich:1970p2023} before being smoothed onto a lower resolution grid.

The ionisation field is then estimated from this evolved IGM density field using the excursion-set approach \citep{Furlanetto:2004p123}. The time-integrated number of ionising photons\footnote{
This includes ionisations from both UV and X-ray sources. While X-rays can produce some level of pre-reionisation \citep[e.g.][]{Ricotti:2004p7145,Dijkstra:2012p7165,McQuinn:2012p3773,Mesinger:2013p1835} their predominant contribution is in pre-heating the IGM before reionisation \cite[e.g.][]{McQuinn:2012p3776}.
} is compared to the number of neutral atoms within regions of decreasing radius, $R$. These regions are computed from a maximum photon horizon, $R_{\rm mfp}$ down to the individual pixel resolution of a single voxel, $R_{\rm cell}$. A voxel at coordinates $(\boldsymbol{x},z)$ within the simulation volume is then tagged as fully ionised if,
\begin{eqnarray} \label{eq:ioncrit}
\zeta f_{\rm coll}(\boldsymbol{x},z,R,\bar{M}_{\rm min}) \geq 1,
\end{eqnarray}
where $f_{\rm coll}(\boldsymbol{x},z,R,\bar{M}_{\rm min})$ is the fraction of collapsed matter residing within haloes more massive than $\bar{M}_{\rm. min}$ \citep{Press:1974p2031,Bond:1991p111,Lacey:1993p115,Sheth:1999p2053} and $\zeta$ is an ionising efficiency describing the conversion of mass into ionising photons (see Section~\ref{sec:Zeta}). Partial ionisations are included for voxels not fully ionised by setting their ionised fractions to
$\zeta f_{\rm coll}(\boldsymbol{x},z,R_{\rm cell},\bar{M}_{\rm min})$.

Since 21 cm observations use the CMB as a background source, the IGM spin temperature has to be coupled to the kinetic gas temperature for the signal to be detected.  This coupling is achieved through either collisional coupling or the  \lya{} background from the first generation of stars (so-called Wouthuysen-Field (WF) coupling \citep{Wouthuysen:1952p4321,Field:1958p1}). To compute the spin temperature, \cmfst{} solves for the evolution of the ionisation, temperature and impinging Ly$\alpha$ background in each voxel\footnote{
Note that within this work we do not vary the soft UV spectra of the first sources driving this epoch. This WF coupling epoch will be extremely challenging to detect in comparison to the EoR and the EoH. Nevertheless, we will return to this in future work. For more specifics regarding the computation of the Ly$\alpha$ background, we refer the reader to Section~3.2 of \citep{Mesinger:2011p1123}}.  These depend on the angle-averaged specific intensity, $J(\textbf{x}, E, z)$, (in erg s$^{-1}$ keV$^{-1}$ cm$^{-2}$ sr$^{-1}$), computed by integrating the comoving X-ray specific emissivity, $\epsilon_{\rm X}(\textbf{x}, E_e, z')$ back along the light-cone:
\begin{equation} \label{eq:Jave}
J(\textbf{x}, E, z) = \frac{(1+z)^3}{4\pi} \int_{z}^{\infty} dz' \frac{c dt}{dz'} \epsilon_{\rm X}  e^{-\tau}.
\end{equation}
Here, $e^{-\tau}$ corresponds to the probability that a photon emitted at an earlier time, $z'$, survives until $z$ owing to IGM attenuation (see Eq.\ 16 of \citealt{Mesinger:2011p1123}) and the comoving specific emissivity is evaluated in the emitted frame, $E_{\rm e} = E(1 + z')/(1 + z)$, with
\begin{equation} \label{eq:emissivity}
\epsilon_{\rm X}(\textbf{x}, E_{\rm e}, z') = \frac{L_{\rm X}}{\rm SFR} \left[ \rho_{\rm crit, 0} \Omega_b f_{\ast} (1+\delta_{\rm nl})\frac{d f_{\rm coll}(z')}{dt} \right],
\end{equation}
where the quantity in square brackets is the star-formation rate (SFR) density along the light-cone, with $\rho_{\rm crit, 0}$ being the current critical density and $f_\ast$ the fraction of galactic baryons converted into stars.

The quantity $L_{\rm X}/{\rm SFR}$ is the specific X-ray luminosity per unit star formation escaping the galaxies (in units of erg s$^{-1}$ keV$^{-1}$ $M^{-1}_{\odot}$ yr). The specific luminosity is taken to be a power law in photon energy, $L_{\rm X} \propto E^{- \alpha_X}$, with photons below some threshold energy, $E_0$, being absorbed inside the host galaxy.\footnote{We note that in the computation of the heating and ionisation rates in the default version of \cmfst{}, these integrals were performed out to infinity, which could result in divergent behaviour for $\alpha_{\rm X} \leq 0$.  Here we adopt an upper limit of 10~keV for computing the rate integrals.  This choice is arbitrary and purely for numerical convenience, as the EoH only depends on the SED below $\sim 2$ keV.  Actual SEDs of course do not diverge but turn over at high energies:  $\sim$ 10 -- 100 keV (e.g. \citealt{Lehmer:2013p7818,Lehmer:2015p7825}).}

Instead of the number of X-ray photons per stellar baryon used in the default version of \cmfst{}, here we normalise the X-ray efficiency in terms of an integrated soft-band ($<2$~keV)
luminosity per SFR (in erg s$^{-1}$ $M^{-1}_{\odot}$ yr):
\begin{equation} \label{eq:normL}
  L_{{\rm X}<2\,{\rm keV}}/{\rm SFR} = \int^{2\,{\rm keV}}_{E_{0}} dE_e ~ L_{\rm X}/{\rm SFR} ~.
\end{equation}
This parametrisation is both physically-motivated (harder photons have mean free paths longer than the Hubble length and so do not contribute to the EoH; e.g. \citealt{McQuinn:2012p3773,Das:2017p7170}), and easier to directly compare with X-ray observations of local star-forming galaxies.

\subsection{The \cmmc{} astrophysical parameter set} \label{sec:description}

In the previous section, we outlined the semi-numerical approach, using \cmfst{}, to obtain the 3D realisations of the 21 cm brightness temperature field. This parameterisation yields six free parameters to be sampled within our Bayesian framework. In this section, we provide more detailed descriptions for each of these parameters, providing physical intuition for their impact on the IGM through X-ray heating or ionisation and defining their allowed ranges.

\begin{figure*} 
	\begin{center}
		\includegraphics[trim = 0cm 0.5cm 0cm 0.5cm, scale = 0.76]{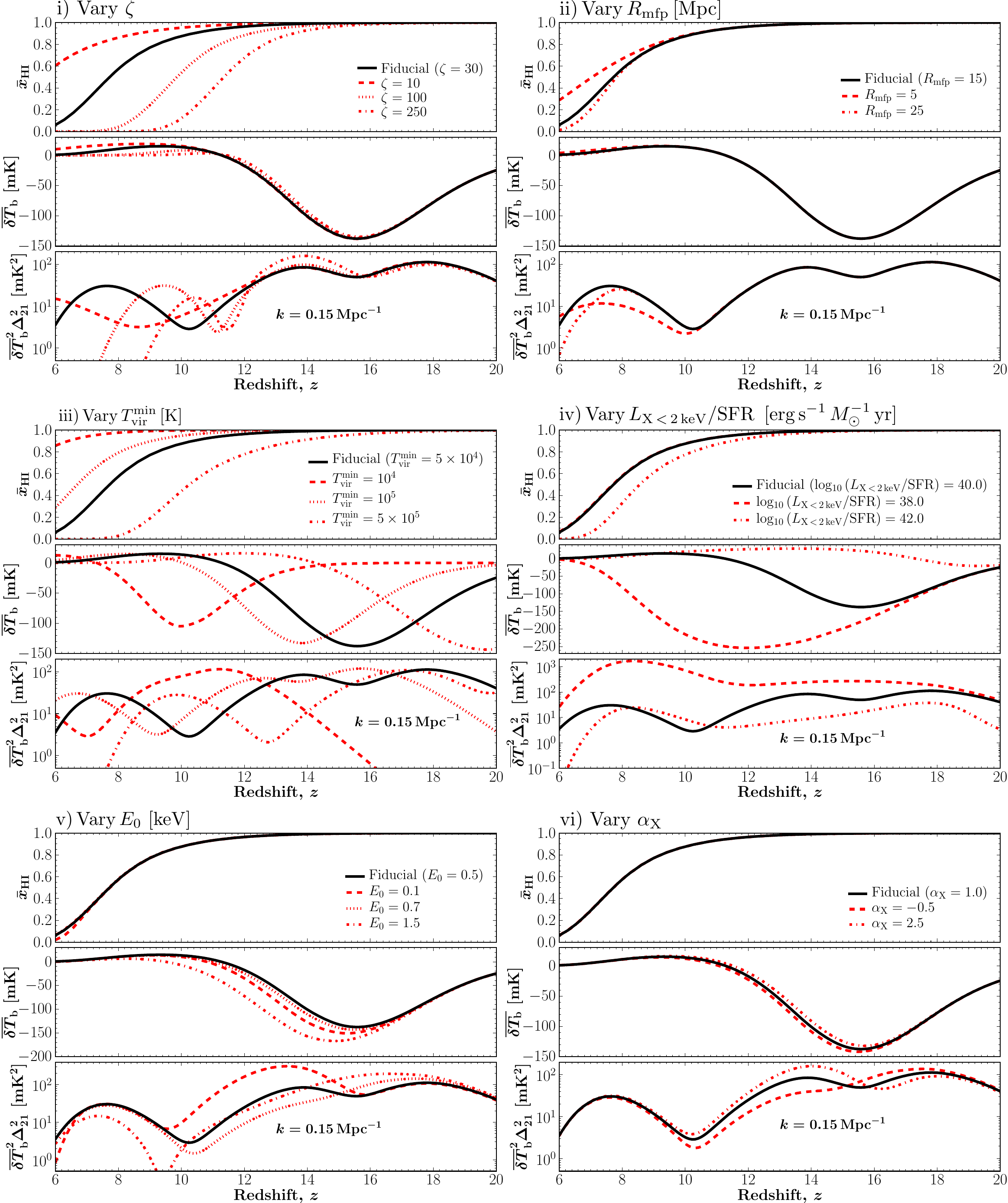}
	\end{center}
\caption[]
{The impact of each of our six astrophysical model parameters on the evolution of the IGM neutral fraction (top panel), global averaged 21 cm brightness temperature contrast (middle panel) and the evolution of the PS amplitude at $k = 0.15$~Mpc$^{-1}$ (bottom panel). We vary each model parameter across the full allowed prior range, holding the remaining five parameters fixed to the fiducial \textsc{faint galaxies} model values (see Section~\ref{sec:MockObs}).}
 \label{fig:ParamVary}
\end{figure*}

To aid in this discussion, we provide Figure~\ref{fig:ParamVary}. For each parameter, denoted (i) - (vi), we provide the redshift evolution of the corresponding average neutral fraction ({\it top panel}),  average 21 cm brightness temperature contrast ({\it middle panel}) and the amplitude of the 21 cm PS at $k = 0.15$~Mpc$^{-1}$ ({\it bottom panel})\footnote{Our choice of $k$-modes was motivated by the fact that the large-scale power in the 21 cm PS is the main discriminant between astrophysical models \citep[e.g.][]{McQuinn:2007p1665, Greig:2015p3675},
  Moreover, these $k$-modes are detectable by upcoming experiments (i.e. they are sufficiently small to avoid being contaminated by foregrounds yet large enough to obtain high signal to noise).
} ($\bar{\delta T^{2}_{\rm b}}\Delta^{2}_{21}(k,z) \equiv k^{3}/(2\pi^{2}V)\,\delta \bar{T}^{2}_{\rm b}(z)\,\langle |\delta_{21}(\boldsymbol{k},z)|^{2}\rangle_{k}$
 where $\delta_{21}(\boldsymbol{x},z) \equiv \delta T_{\rm b}(\boldsymbol{x},z)/\bar{\delta T_{\rm b}}(z) -1$). In each set of panels, we vary the astrophysical parameter in question across its full allowed span, holding the other five parameters fixed.

The solid, black curve in each panel corresponds to our `fiducial' \textsc{faint galaxies} model (see Section~\ref{sec:MockObs} for further details).  Before discussing the impact of each parameter, it is instructive to note the general features exhibited by the fiducial model.  The EoR history in the top panel has a monotonic evolution with a midpoint around $z\sim8$.  The global 21 cm signal in the middle panel shows a deep absorption trough, corresponding to when the X-ray heating rate surpasses the adiabatic cooling rate (start of the EoH). This is followed by a small emission peak, corresponding to the onset of the EoR \citep[e.g.][]{Furlanetto:2006p3782}. The height of the peak is determined by the relative overlap of the EoR and EoH; if the overlap is strong, reionisation can proceed in a cold IGM, with the global signal never switching to emission.

The large-scale power in the bottom panel shows the characteristic three-peaked structure, driven by fluctuations in the \lya{} background (WF coupling); IGM gas temperature (EoH), and the ionisation fraction (EoR), from right to left, respectively \citep[e.g.][]{Pritchard:2007p3787,Baek:2010p6357}. The redshift position of each peak therefore traces the timing of each epoch, while the amplitude of the 21 cm power traces the level of fluctuations determined by the typical source bias and/or X-ray SED \citep[e.g.][]{Mesinger:2013p1835}. Troughs on the other hand correspond to the transitions between epochs, when the corresponding cross terms of the 21 cm PS cause the large-scale power to drop \citep[e.g.][]{Lidz:2007p1929,Pritchard:2007p3787,Mesinger:2016p6167}.
 
\subsubsection{UV ionising efficiency, $\zeta$} \label{sec:Zeta}

The UV ionising efficiency of high-$z$ galaxies (Equation~\ref{eq:ioncrit}) can be expressed as 
 \begin{eqnarray} \label{eq:Zeta}
\zeta = 30\left(\frac{f_{\rm esc}}{0.12}\right)\left(\frac{f_{\ast}}{0.05}\right) \left(\frac{N_{\gamma/b}}{4000}\right)\left(\frac{1.5}{1+n_{\rm rec}}\right)
 \end{eqnarray}
 where, $f_{\rm esc}$ is the fraction of ionising photons escaping into the IGM, $f_{\ast}$ is the fraction of galactic gas in stars, $N_{\gamma/b}$ is the number of ionising photons produced per baryon in stars and $n_{\rm rec}$ is the typical number of times a hydrogen atom recombines. While only the product of equation~(\ref{eq:Zeta}) is required for generating the ionisation field, we provide some plausible values for each of the terms. We adopt $N_{\gamma}\approx4000$ as is typically expected from Population II stars \citep[e.g.][]{Barkana:2005p1934}. For high-z galaxies, both $f_{\star}$ and $f_{\rm esc}$ are extremely uncertain, with observations and simulations placing plausible values within the vicinity of $\sim10$ per cent for $f_{\star}$ \citep[e.g.][]{Behroozi:2015p1,Sun:2016p8225} and $f_{\rm esc}$ \citep[e.g][]{Paardekooper:2015p1,Xu:2016p1,Kimm:2017p7875}. Finally, we adopt $n_{\rm rec} \sim 0.5$, similar to those found in the 'photon-starved' reionisation models of \citep{Sobacchi:2014p1157}, which are consistent with emissivity estimates from the \lya{} forest \citep[e.g.][]{Bolton:2007p3273,McQuinn:2011p3293}. It is important to note that the $f_{\ast}$ appearing here, is equivalent to the $f_{\ast}$ in equation~(\ref{eq:emissivity}). That is, throughout all cosmic epochs, we adopt the same, constant value of 5 per cent for the fraction of galactic gas in stars. In future work, we will relax this assumption.
    
In this work, we adopt a flat prior over the fiducial range of $\zeta\in[10,250]$, which is a notable extension over the original upper limit of $\zeta=100$ chosen in \citet{Greig:2015p3675}. The extended range provides the flexibility to explore physically plausible models where reionisation is driven by bright, rare galaxies (see Section~\ref{sec:MockObs}). In panel (i) of Figure~\ref{fig:ParamVary} we highlight the impact of varying $\zeta$. As expected, $\zeta$ has a strong impact on the EoR and an almost negligible impact on the EoH. As $\zeta$ is increased, the EoR peak shifts to earlier redshifts and the width of the EoR peak reduces (i.e. shorter duration). For extremely large values of $\zeta$, the EoR and EoH peaks begin to merge, resulting in a larger amplitude EoR peak, sourced by the contrast between the cold IGM patches present in the early EoH stages and the \hii{} regions.
 
\subsubsection{Maximum ionising photon horizon within ionised regions, $R_{\rm mfp}$} 

The physical size of \hii{} regions is regulated by the distance ionising photons can propagate into the IGM. This depends on the abundance of photon sinks (absorption systems such as Lyman limit systems) and the corresponding recombinations of these systems.  When the \hii{} regions start approaching the typical separation of the photon sinks, an increasing fraction of ionising photons are used to balance recombinations, and the EoR can slow down \citep[e.g.][]{Furlanetto:2005p4326,Furlanetto:2009p7171,Alvarez:2012p1930}.

The details of this process can be complicated (e.g. \citealt{Sobacchi:2014p1157}); however a common simplification in semi-numerical approaches is to adopt a maximum horizon for the ionising photons within the ionised IGM, $R_{\rm mfp}$\footnote{
  For historical context, we adopt `$R_{\rm mfp}$' to denote this effective horizon set by sub-grid recombinations. The sub-grid models of \citet{Sobacchi:2014p1157} do not translate to a constant, uniform value of $R_{\rm mfp}$, and further the mean free path is typically larger than $R_{\rm mfp}$ (since the former is an instantaneous quantity while the latter depends on the cumulative contributions of sub-grid recombinations). However, these authors find that the approximation of a constant $R_{\rm mfp}$ can reproduce the 21 cm PS of their fiducial model to $\sim10$ per cent. Future versions of \cmmc{} will include the \citet{Sobacchi:2014p1157} sub-grid recombination model, eliminating $R_{\rm mfp}$ as a free parameter.
},
which is implemented as the maximum filtering scale in the excursion-set EoR modelling (see Section~\ref{sec:modelling}). Motivated by recent sub-grid recombination models \citep{Sobacchi:2014p1157}, we adopt a flat prior over $R_{\rm mfp}\in[5, 25]$~cMpc. 
 
In panel (ii) of Figure~\ref{fig:ParamVary}, we show the impact of $R_{\rm mfp}$ over cosmic history.  As mentioned above, $R_{\rm mfp}$ only becomes important in the advanced stages of reionization, when the typical \hii{} region scale approaches $R_{\rm mfp}$.  The result is a delay of the late stages of the EoR for small values of $R_{\rm mfp}$. Limiting the photon horizon with a decreasing $R_{\rm mfp}$ ($\lesssim15$~Mpc) is also evidenced by a drop in the large-scale 21 cm PS \citep[e.g.][]{McQuinn:2007p1665, Alvarez:2012p1930,Mesinger:2012p1131,Greig:2015p3675}. Values of  $R_{\rm mfp}>15$~Mpc have little impact on the 21 cm PS (for this combination of the other astrophysical parameters), as the clustering of the ionising sources becomes the dominant source of power (note the same PS amplitude for $R_{\rm mfp} = 15$ and 25~Mpc).
  
\subsubsection{Minimum virial temperature of star-forming haloes, $T^{\rm min}_{\rm vir}$} \label{sec:Tvir}

We define the minimum threshold for a halo hosting a star-forming galaxy to be its virial temperature, $T^{\rm min}_{\rm vir}$, which is related to the halo mass via, \citep[e.g.][]{Barkana:2001p1634}
\begin{eqnarray}
M^{\rm min}_{\rm vir} &=& 10^{8} h^{-1} \left(\frac{\mu}{0.6}\right)^{-3/2}\left(\frac{\Omega_{\rm m}}{\Omega^{z}_{\rm m}}
\frac{\Delta_{\rm c}}{18\pi^{2}}\right)^{-1/2} \nonumber \\
& & \times \left(\frac{T^{\rm min}_{\rm vir}}{1.98\times10^{4}~{\rm K}}\right)^{3/2}\left(\frac{1+z}{10}\right)^{-3/2}M_{\sun},
\end{eqnarray}
where $\mu$ is the mean molecular weight, $\Omega^{z}_{\rm m} = \Omega_{\rm m}(1+z)^{3}/[\Omega_{\rm m}(1+z)^{3} + \Omega_{\Lambda}]$, and $\Delta_{c} = 18\pi^{2} + 82d - 39d^{2}$ where $d = \Omega^{z}_{\rm m}-1$.  The choice of $T^{\rm min}_{\rm vir}$ acts as a step-function cut-off to the UV luminosity function. Below $T^{\rm min}_{\rm vir}$, it is assumed that internal feedback mechanisms such as supernova or photo-heating  suppress the formation of stars. Above, efficient star formation overcomes internal feedback, enabling these haloes to produce ionising photons capable of contributing to the EoR and EoH. We shall consider a flat prior across $T^{\rm min}_{\rm vir}\in[10^{4},10^{6}]$~K within this work.

Our lower limit, $T^{\rm min}_{\rm vir}\approx10^{4}$~K is motivated by the minimum temperature for efficient atomic line cooling. In principle, $T^{\rm min}_{\rm vir}$ can be as low as $\approx10^{2}$~K in the presence of radiative cooling \citep{Haiman:1996p2144,Tegmark:1997p7180,Abel:2002p2149,Bromm:2002p2153}, however, star formation within these haloes is likely inefficient (a few stars per halo; e.g.~\citealt{Kimm:2017p7875}) and can quickly ($z>20$) be suppressed by Lyman-Werner or other feedback processes well before the EoR \citep{Haiman:2000p2155,Ricotti:2001p2160,Haiman:2006p2169,Mesinger:2006p2171,Holzbauer:2012p2890,Fialkov:2013p2903}. Our upper limit of $T^{\rm min}_{\rm vir}\approx10^{6}$~K, is roughly consistent with the host halo masses of observed Lyman break galaxies at $z\sim6$--8, as estimated with the abundance matching technique (e.g. \citealt{Kuhlen:2012p1506, BaroneNugent:2014p4324}).

Note that within the \cmfst{} framework, $T^{\rm min}_{\rm vir}$ is important both in the EoR and EoH, determining the ionisation field ($f_{\rm coll}$, equation~\ref{eq:ioncrit}) and the specific emissivity requisite for the X-ray heating background ($\frac{df_{\rm coll}}{dt}$, equation~\ref{eq:emissivity}) respectively. This implies that the efficient star-forming galaxies responsible for reionisation are the same galaxies which host the sources responsible for X-ray heating (i.e. the physics of star formation drives both the X-ray heating and ionisation fields). As a result, $T^{\rm min}_{\rm vir}$ affects the timing of both the EoR and EoH.

This is evident in panel (iii) of Figure~\ref{fig:ParamVary}.  The EoR and EoH milestones are pushed to lower redshifts for an increasing $T^{\rm min}_{\rm vir}$. For a suitably large choice of $T^{\rm min}_{\rm vir}$ (e.g. $5\times10^5$~K), the timing of the EoH and WF-coupling peaks can be sufficiently delayed to result in their overlap. In addition to the timing, $T^{\rm min}_{\rm vir}$, also impacts the amplitude of the fluctuations.  Rarer (more biased) galaxies, corresponding to larger values of  $T^{\rm min}_{\rm vir}$, are evidenced by more 21 cm power at a given epoch \citep[e.g.][]{McQuinn:2007p1665}.

\subsubsection{Integrated soft-band luminosity, $L_{\rm X\,<\,2\,keV}/{\rm SFR}$} \label{sec:LX}

The efficiency of X-ray heating is driven by the total integrated soft-band ($<$~2~keV) luminosity per SFR (equation~\ref{eq:normL}) escaping into the IGM, which normalises the emergent specific emissivity produced by the X-ray sources within the first galaxies.
Within this work, we adopt a flat prior over the range ${\rm log}_{10}(L_{\rm X\,<\,2\,keV}/{\rm SFR})\in[38,42]$. This range is conservatively selected\footnote{
It is also roughly consistent with the limits proposed by \citet{Fialkov:2017p6998} using observations of the unresolved cosmic X-ray background \citep[e.g.][]{Lehmer:2012p7191} and the upper limits on the measured 21 cm PS from PAPER-64 \citep[e.g.][]{Ali:2015p4327,Pober:2015p4328}.
} to be one to two orders of magnitude broader than the distribution seen in local populations of star forming galaxies \citep{Mineo:2012p6282}, and their stacked {\it Chandra} observations \citep{Lehmer:2016p7810}.  It also encompasses values at high-redshifts predicted by population synthesis models \citep{Fragos:2013p6529}.

In Panel (iv) of Figure~\ref{fig:ParamVary} we highlight the impact of $L_{\rm X\,<\,2\,keV}/{\rm SFR}$. For very high values of $L_{\rm X\,<\,2\,keV}/{\rm SFR}$, the EoH commences prior to the completion of WF-coupling. As a result, no strong absorption feature in $\bar{\delta T_{\rm b}}$ is observed, and the \lya{}-EoH peaks in the 21 cm PS merge.  In addition to heating, such high X-ray luminosities can also substantially ionise the EoR (at the $\sim10$-20 per cent level).  In this case, the EoR can complete earlier (top panel).

At the other end of the range,  extremely in-efficient X-ray heating (dashed curve) results in a delayed EoH, with the EoR and EoH overlapping.  In this `cold reionisation' scenario \citep{Mesinger:2013p1835}, reionisation proceeds in an IGM which is significantly colder than the CMB.
The resultant large temperature contrasts between the ionised \hii{} regions and cold neutral IGM can yield an extremely large 21 cm PS amplitude ($\sim10^3$ mK$^{2}$).\footnote{Though the specific model shown here exceeds the PAPER-64 upper limits, for other choices of the EoR and EoH parameters, this same choice of $L_{\rm X\,<\,2\,keV}/{\rm SFR}=10^{38}$~erg~s$^{-1}\,M^{-1}_{\odot}$~yr can result in a 21 cm PS below these upper limits.}

\subsubsection{X-ray energy threshold for self-absorption by the host galaxies, $E_{0}$} \label{sec:E0}

The soft X-rays produced within galaxies can be absorbed by the intervening ISM, thus not being able to escape and contribute to heating the IGM.
The impact that this ISM absorption has on the emergent X-ray SED depends on the ISM density and metallicity. Early galaxies responsible for the EoH are expected to be less polluted by metals than local analogues. Indeed, \citet{Das:2017p7170} find that the emergent X-ray SED from simulated high-redshift galaxies can be well approximated by a metal free ISM with a typical column density of ${\rm log_{10}}(N_{\rm \hi{}}/{\rm cm^{2}}) = 21.40\substack{+0.40 \\ -0.65}$. As the opacity of metal free gas is a steep function of energy, these authors find that the 21 cm PS from the emerging X-ray SED can be well approximated using the common assumption of a step function attenuation of the X-ray SED below an energy threshold, $E_{0}$. In this work, we adopt a flat prior over $E_{0}\in[0.1,1.5]$~keV which corresponds to ${\rm log_{10}}(N_{\rm \hi{}}/{\rm cm^{2}})\in[19.3,23.0]$\footnote{The conversion to column densities is computed assuming a unity optical depth for a pristine, metal free, neutral ISM (i.e. the contribution from \heii{} is negligible).}.

Panel (v) of Figure~\ref{fig:ParamVary} highlights the impact of $E_{0}$. As $L_{\rm X\,<\,2\,keV}/{\rm SFR}$ defines the total soft-band luminosity, as we increase (decrease) $E_{0}$ we effectively harden (soften) the spectrum of emergent X-ray photons.
Since the absorption cross section scales as $\propto E^{-3}$, we would naively expect smaller values of $E_{0}$ to result in more efficient heating, shifting the minimum in the global signal to higher redshifts.  However, this evolution is slightly reversed for very low values, $E_{0} \lsim 0.5$ keV, since the energy of these very soft photons is continually deposited in a limited volume surrounding the first galaxies; thus the volume-averaged global signal during the EoH is slightly delayed for these very soft SEDs.  This highlights the need to properly model the spatially-dependent IGM heating, even when predicting the average signal.

The amplitude of the PS is very highly dependent on $E_0$.  Softer SEDs result in very inhomogeneous heating, with PS amplitudes larger by up to an order of magnitude (e.g. \citealt{Pacucci:2014p4323}).
The amplitude of the EoH peak consistently decreases for an increasing $E_{0}$, as the harder SEDs make the EoH more homogeneous.  Eventually for $E_{0} > 0.7$~keV, no EoH peak occurs, as the large mean free path of the X-ray photons result in an inefficient, relatively uniform heating of the IGM \citep[e.g.][]{Fialkov:2014p3720}.
  
\subsubsection{X-ray spectral index, $\alpha_{\rm X}$} \label{sec:AlphaX}

The spectral index, $\alpha_{\rm X}$, describing the emergent spectrum from the first galaxies hosting X-ray sources depends on what is assumed to be the dominant process producing X-ray photons. In this work, we adopt a fiducial flat prior of $\alpha_{\rm X}\in[-0.5,2.5]$ which should encompass the most relevant high energy X-ray SEDs that describe the first galaxies (e.g. HMXBs, host ISM, mini-quasars, SNe remnants etc.; see for example \citealt{McQuinn:2012p3773,Pacucci:2014p4323}).

Finally, in panel (vi) of Figure~\ref{fig:ParamVary}, we illustrate the impact of $\alpha_{\rm X}$.
For an increasing $\alpha_{\rm X}$, more soft X-ray photons are produced (as the soft-band luminosity is kept fixed) resulting in more efficient X-ray heating. This results in an increase in the temperature fluctuations, driving a larger amplitude of the EoH peak in the 21 cm PS. Conversely, for an inverted spectral index, $\alpha_{\rm X} = -0.5$, the emergent X-ray photons are spectrally harder, producing inefficient uniform X-ray heating, and limited temperature fluctuations. Thus, the EoH peak in the 21 cm PS for $\alpha_{\rm X} = -0.5$ (dashed curve) is suppressed.  However, over our adopted range of priors, the absorption by the host galaxy, $E_{0}$, is much more potent in hardening/softening the SED than the spectral index, $\alpha_{\rm X}$.

\subsection{Telescope noise profiles} \label{sec:noise}

In order to provide astrophysical parameter forecasting, we must model the expected noise of the 21 cm experiments. Within this work, we focus solely on the 21 cm PS. To generate the sensitivity curves for 21 cm interferometer experiments, we use the python module \sense{}\footnote{https://github.com/jpober/21cmSense}\citep{Pober:2013p41,Pober:2014p35}. Below, we summarise the main aspects and assumptions required to produce telescope noise profiles.

Firstly, the thermal noise PS is calculated at each gridded $uv$-cell according to the following \citep[e.g.][]{Morales:2005p1474,McQuinn:2006p109,Pober:2014p35},
\begin{eqnarray} \label{eq:NoisePS}
\Delta^{2}_{\rm N}(k) \approx X^{2}Y\frac{k^{3}}{2\pi^{2}}\frac{\Omega^{\prime}}{2t}T^{2}_{\rm sys},
\end{eqnarray} 
where $X^{2}Y$ is a cosmological conversion factor between the observing bandwidth, frequency and comoving distance, $\Omega^{\prime}$ is a beam-dependent factor derived in \citet{Parsons:2014p781}, $t$ is the total time spent by all baselines within a particular $k$-mode and $T_{\rm sys}$ is the system temperature, the sum of the receiver temperature, $T_{\rm rec}$, and the sky temperature $T_{\rm sky}$. We model $T_{\rm sky}$ using the frequency dependent scaling $T_{\rm sky} = 60\left(\frac{\nu}{300~{\rm MHz}}\right)^{-2.55}~{\rm K}$ \citep{Thompson2007}.

The sample variance of the cosmological 21 cm PS can easily be combined with the thermal noise to produce the total noise PS using an inverse-weighted summation over all the individual modes \citep{Pober:2013p41},
\begin{eqnarray} \label{eq:T+S}
\delta\Delta^{2}_{\rm T+S}(k) = \left(\sum_{i}\frac{1}{(\Delta^{2}_{{\rm N},i}(k) + \Delta^{2}_{21}(k))^{2}}\right)^{-\frac{1}{2}},
\end{eqnarray}
where $\delta\Delta^{2}_{\rm T+S}(k)$ is the total uncertainty from thermal noise and sample variance in a given $k$-mode and $\Delta^{2}_{21}(k)$ is the cosmological 21 cm PS (mock observation). Here we assume Gaussian errors for the cosmic-variance term, which is a good approximation on large-scales.

The largest primary uncertainty for 21 cm experiments is dealing with the bright foreground contamination. However, for the most part these bright foregrounds are spectrally smooth and have been shown to reside within a confined region of cylindrical 2D $k$-space known as the `wedge' \citep{Datta:2010p2792,Vedantham:2012p2801,Morales:2012p2828,Parsons:2012p2833,Trott:2012p2834,Hazelton:2013p1481,Thyagarajan:2013p2851,Liu:2014p3465,Liu:2014p3466,Thyagarajan:2015p7294,Thyagarajan:2015p7298,Pober:2016p7301}. Outside of this `wedge' we are left with a relatively pristine observing window where the cosmic 21 cm signal is only affected by the instrumental thermal noise. At present, the location of the boundary separating this observing window and the `wedge'-like feature is uncertain.

Within \sense{} three foreground removal strategies are provided \citep{Pober:2014p35}, ``optimistic'', ``moderate'' and ``pessimistic''. We defer the reader to this work for further details on these scenarios, highlighting here that we choose to adopt the ``moderate'' scenario. This entails 21 cm observations only within the pristine 21 cm window (i.e. avoiding the ``wedge''), with the wedge location defined to extend $\Delta k_{\parallel} = 0.1 \,h$~Mpc$^{-1}$ beyond the horizon limit \citep{Pober:2014p35}. Furthermore, this scenario includes the coherent summation over all redundant baselines within the array configuration allowing the reduction of thermal noise \citep{Parsons:2012p95}.

\begin{table}
\begin{tabular}{@{}lcccc}
\hline
Parameter & HERA & SKA \\
\hline
Telescope antennae  & 331 & 512 \\
Diameter (m)  & 14 & 35 \\
Collecting Area (m$^2$)  & 50\,953 & 492\,602\\
$T_{\rm rec}$ (K)  & 100 & 0.1T$_{\rm sky}$ + 40 \\
Bandwidth (MHz)  & 8 & 8 \\
Integration time (hrs) & 1000 (drift) & 1000 (tracked) \\
\hline
\end{tabular}
\caption{Summary of telescope parameters we use to compute sensitivity profiles (see text for further details).}
\label{tab:TelescopeParams}
\end{table}

Within this work, we focus on the two second generation 21 cm interferometer experiments capable of simultaneously measuring the EoR and the EoH, namely the SKA and HERA. Below (and in Table~\ref{tab:TelescopeParams}) we summarise the specific design features and assumptions required to model the theoretical noise of both instruments:

\begin{itemize}
\item[1.] \textit{HERA:} We follow the design specifics outlined in \citet{Beardsley:2014p1529} with a core design consisting of 331 dishes\footnote{Note, the final fully funded design for HERA consists of 350 dishes, 320 in the core and 30 outriggers \citep{DeBoer:2017p6740}. For all intents and purpose for this work, the difference between a 320 and 331 core layout is negligible.}. Each dish is 14m in diameter closely packed into a hexagonal configuration to maximise the total number of redundant baselines \citep{Parsons:2012p95}. We model the total system temperature as $T_{\rm sys} = 100 + T_{\rm sky}~{\rm K}$. HERA will operate in a drift-scanning mode, for which we assume a total 1080~hr observation, spread across 180 nights at 6 hours per night.
 \\
\item[2.] \textit{SKA:} We use the latest\footnote{
In \citet{Greig:2015p3675} we used the original design prior to the re-baselining (the 50 per cent reduction in the number of antennae dipoles), while in \citet{Greig:2015p6161} we investigated several design layouts to maximise the sensitivity specifically for the 21 cm PS following the re-baselining. The latest design for the SKA results in reduced sensitivity when compared to both our previous works, therefore we caution comparisons between the SKA within this work and our previous studies.
} available design for SKA-low Phase 1, using the telescope positions provided in the most recent SKA System Baseline Design document\footnote{http://astronomers.skatelescope.org/wp-content/uploads/2016/09/SKA-TEL-SKO-0000422\textunderscore 02\textunderscore SKA1\textunderscore LowConfigurationCoordinates-1.pdf}. Specifically, SKA-low Phase 1 includes a total of 512 35m antennae stations randomly distributed within a 500m core radius.  The total SKA system temperature is modelled as outlined in the SKA System Baseline Design, $T_{\rm sys} = 1.1T_{\rm sky} + 40~{\rm K}$. For the SKA we adopt a single, deep 1000~hr tracked scan of a cold patch on the sky.
\end{itemize}

It is non trivial to perform a like-for-like comparison between the two experiments, as HERA intends to perform a rotational synthesis drift scan, whereas the SKA intends to conduct track scanned observations\footnote{Additionally, HERA is a dedicated 21 cm experiment specifically designed for a 21 cm PS measurement, while the SKA is a multidisciplinary experiment, with detection of the 21 cm signal only one of many key science goals. Moreover, the instrument layout and design is tailored towards 3D tomographic imaging \citep[e.g.][]{Mellema:2013p2975} rather than the 21 cm PS.}. These two strategies result in considerably different noise PS. A single tracked field with the SKA will have considerably lower thermal noise than HERA owing to the deeper integration time, therefore SKA will be superior on small scales (large $k$) important for imaging. On the other hand, by observing numerous patches of the sky rotating through the zenith pointing field of view of HERA per night, sample variance can be better mitigated compared to the single tracked field of the SKA (i.e. HERA will have reduced noise on large scales, small $k$). For the most part, the strongest constraints on the astrophysical parameters come from the large-scale (small $k$) modes of the 21 cm PS, therefore using the PS as the likelihood statistic will favour the approach of HERA over that of SKA.  We quantify these claims further in the Appendix, showing the noise power spectra at different redshifts.

Further complicating a direct comparison is the fact that the SKA is planning a tiered survey.
For simplicity, in this introductory work we only consider a single, deep 1000~hr observation (though we will return to this in the future). However, we could have considered either the intermediate $10\times100$~hr or wide and shallow $100\times10$~hr strategies \citep[e.g.][]{Greig:2015p6161}. These latter two surveys concede thermal noise sensitivity (from the reduced per field integration time) for an increased sample variance sensitivity by surveying multiple fields. In terms of the simplified three parameter EoR model considered in \citet{Greig:2015p3675}, the single, deep 1000~hr observation recovered the largest uncertainties on the astrophysical parameters relative to the intermediate or wide and shallow surveys (i.e. a single, tracked field was the worst performed strategy).

\section{Mock cosmic signal} \label{sec:MockObs} 

Having outlined our astrophysical model to describe the EoR and EoH, we now introduce our mock observations of the cosmic 21 cm signal. It is impractical to vary all available astrophysical parameters when creating mock observations, therefore, following \citet{Mesinger:2016p6167} we take two extreme choices for $T^{\rm min}_{\rm vir}$. This parameter characterises both the timing of the epochs and the typical bias of the dominant galaxies, thus encoding the largest variation in the 21 cm signal (e.g panel (iii) of Figure~\ref{fig:ParamVary}). Specifically, we adopt $T^{\rm min}_{\rm vir} = 5\times10^4$~K (\textsc{faint galaxies}) and $T^{\rm min}_{\rm vir} = 3\times10^5$~K (\textsc{bright galaxies}). These choices approximately match the evolution of the cosmic SFR density inferred from extrapolating the observed luminosity functions\footnote{
In the future, we will provide a more generalised parameterisation of the ionising source model by allowing each individual constituent of $\zeta$ (equation~\ref{eq:Zeta}) to vary with halo mass (Park et al., in prep.). This will enable additional flexibility in the source modelling, while at the same time allowing high-$z$ galaxy luminosity functions to be applied as priors to the source model prescription.
} \citep{Bouwens:2015p7832} down to a UV magnitude of $M_{\rm UV} = -10 \,(-17)$ for the \textsc{faint galaxies} (\textsc{bright galaxies}) galaxies model (see figure~5 of \citealt{Das:2017p7170}).

In order to match the latest constraints on the electron scattering optical depth, $\tau_{\rm e}$, from Planck ($\tau_{\rm e} = 0.058 \pm 0.012$; \citealt{PlanckCollaboration:2016p7627}), we adopt $\zeta = 30$ for the \textsc{faint galaxies} model. This corresponds to the fiducial parameter set outlined in equation~(\ref{eq:Zeta}). For the biased, rarer galaxies in the \textsc{bright galaxies} model, we adopt $\zeta = 200$. We select a fiducial $R_{\rm mfp} = 15$~Mpc for both models. In Table~\ref{tab:ParamSum} we summarise the adopted astrophysical parameter set for each of the two models, while also providing the corresponding optical depth, $\tau_{\rm e}$.

\begin{table*}
\begin{tabular}{@{}lcccccccccc}
\hline
 & $\zeta$ & $R_{\rm mfp}$& $T^{\rm min}_{\rm vir}$ & log$_{10}$$(T^{\rm min}_{\rm vir})$ & log$_{10}$($L_{\rm X\,<\,2\,keV}/{\rm SFR}$) & $E_{0}$ & ${\rm log_{10}}(N_{\rm \hi{}})$ & $\alpha_{X}$ & $\tau_{\rm e}$ \\
Source Model &  & [Mpc] & [K] & [K] & [erg s$^{-1}$ $M^{-1}_{\odot}$ yr] & [keV] & [cm$^{-2}$] & \\
\hline
\vspace{0.8mm}
\textsc{faint galaxies} & 30.0 & 15.0 & $5\times10^{4}$ & 4.70 & 40.0 & 0.5 & 21.45 & 1.0 & 0.059 \\
\vspace{0.8mm}
\textsc{bright galaxies} & 200.0 & 15.0 & $3\times10^{5}$ & 5.48 & 40.0 & 0.5 & 21.45 & 1.0 & 0.056\\
\hline
\vspace{0.8mm}
Allowed range & [10, 250] & [5, 25] & [$10^{4}$, $10^{6}$] & [4.0, 6.0] & [38.0, 42.0] & [0.1, 1.5] & [19.3,23.0] & [-0.5, 2.5] \\
\hline
\end{tabular}
\caption{A summary of the astrophysical parameters used for our two mock observations, the corresponding electron scattering optical depth, $\tau_{\rm e}$, as well as the adopted prior range for \cmmc{}. See text for additional details.
}
\label{tab:ParamSum}
\end{table*}

We adopt the same X-ray source model for both the \textsc{bright galaxies} and \textsc{faint galaxies} models. We assume HMXBs to be the dominant X-ray heating source within the first galaxies, and use the results of \citet{Das:2017p7170} to describe the emergent X-ray SED\footnote{
Note that the typical star-forming haloes used by \citet{Das:2017p7170}, e.g.~$\lesssim3\times10^{8}\,M_{\odot}$, are smaller than the corresponding masses set by our minimum on $T^{\rm min}_{\rm vir}$ (e.g.~$10^4$~K). Therefore, we have assumed that the intrinsic attenuation is independent of the host galaxy mass. This, however, need not be the case. 
}. This corresponds to an energy threshold, $E_{0} = 0.5$~keV (${\rm log_{10}}(N_{\rm \hi{}}/{\rm cm^{2}}) = 21.5$) and an X-ray spectral index of $\alpha_{X} = 1.0$. Finally, we assume a HMXB soft band luminosity of $L_{\rm X\,<\,2\,keV}/{\rm SFR} = 10^{40}$~erg~s$^{-1}\,M^{-1}_{\odot}$~yr, which is consistent with estimates from the HMXB population synthesis models of \citet{Fragos:2013p6529}.

\begin{figure} 
	\begin{center}
		\includegraphics[trim = 0cm 0.9cm 0cm 0.5cm, scale = 0.415]{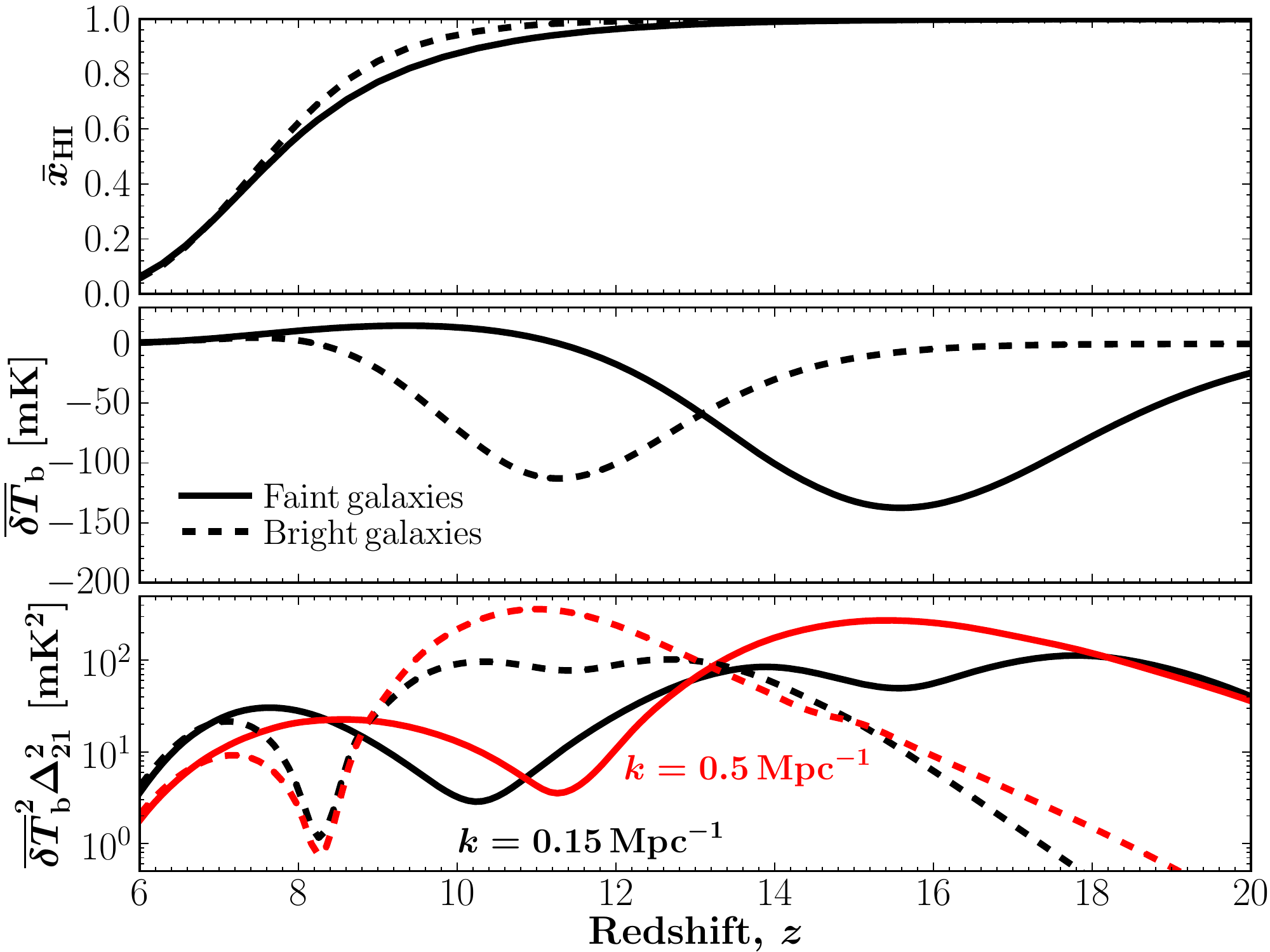}
	\end{center}
\caption[]
{Cosmic evolution of the \textsc{bright galaxies} (dashed) and \textsc{faint galaxies} (solid) models used for the mock observations within this work. \textit{Top:} the evolution of the IGM neutral fraction. \textit{Middle:} global averaged 21 cm brightness temperature contrast. \textit{Bottom:} evolution of the PS amplitude at two different $k$-modes, $k = 0.15$~Mpc$^{-1}$ (black) and $k = 0.5$~Mpc$^{-1}$ (red).}
\label{fig:MockObservations}
\end{figure}

In Figure~\ref{fig:MockObservations} we provide the cosmic evolution of our two EoR source models: \textsc{faint galaxies} (solid curve) and \textsc{bright galaxies} (dashed curve). The top, middle and bottom panels are the global reionisation history, global 21 cm signal, and the 21 cm PS at $k=0.15$ (black) and $k=0.5$~Mpc$^{-1}$ (red) respectively. These show qualitatively the same behaviour as we have discussed in the previous section. These mock observations are generated from simulations with a volume of 600$^{3}$ Mpc$^{3}$, on a 400$^3$ grid smoothed down from the high-resolution initial conditions generated on a 2400$^3$ grid.

By construction, these two EoR models have a similar reionisation history, consistent with the latest observations. The \textsc{faint galaxies} model has a somewhat more extended EoR, as the less biased DM halos which host the dominant galaxies form slower.  Unlike for the observationally-constrained EoR history, we allow the EoH history to be different in the two models.  In other words, we take the same X-ray luminosity per SFR.  As a result, the \textsc{faint galaxies} galaxy model has an earlier EoH (see the middle panel), governed by the formation of the first $T^{\rm min}_{\rm vir} = 5\times10^4$~K structures.

Due to this, these two models will be interesting for our astrophysical parameter forecasting for both HERA and the SKA. With a decreasing sensitivity with increasing redshift, the location of the EoR and EoH peaks will impact the resultant significance of a detection of the 21 cm signal. For example, in the bottom panel of Figure~\ref{fig:MockObservations}, it is evident that the \textsc{bright galaxies} model exhibits additional 21 cm power relative to the \textsc{faint galaxies} model. Furthermore, this occurs at a lower redshift, where the instrumental noise is expected to be lower.  Naively, one would therefore expect astrophysical parameters to be more tightly constrained with a mock \textsc{bright galaxies} signal.

\section{21 cm forecasts} \label{sec:Forecasts}

We now quantify astrophysical parameter constraints for each of these two models, for both second generation interferometers, HERA and the SKA. We first summarise our \cmmc{} configuration, and then provide parameter constraints while discussing their implications.

\subsection{\cmmc{} setup}

As in \citet{Greig:2015p3675}, we use the 21 cm PS as the likelihood statistic to sample the astrophysical parameter space in \cmmc{}. We adopt a modelling uncertainty of 20 per cent on the sampled 21 cm PS (not the mock observation of the 21 cm PS), which is added in quadrature with the total noise power spectrum (equation~\ref{eq:NoisePS}).\footnote{A modelling uncertainty accounts for the inaccuracy of semi-numerical approaches such as \cmfst{}.
Here we simply take an uncorrelated fixed percentage error, with a value consistent with the comparisons in \citet{Zahn:2011p1171}. This constant error purely accounts for the observed fractional differences in the recovered 21cm PS between the simulations. For example, it does not account for any potential errors arising from the different algorithms used to compute the IGM spin temperature. Future work will seek to characterise this uncertainty, potentially even mediating it by comparing \cmfst{} to suites of RT simulations.}
As in \citet{Greig:2015p3675} we compare the 21 cm PS at each redshift only over the reduced $k$-space range, $k = 0.15\,$--$\,1.0$~Mpc$^{-1}$, which we deem to be free of cosmic variance and shot-noise effects arising within the simulations.

We perform our forecasting using eight co-evolution\footnote{The observed 3D 21 cm signal is spatially dependent on the 2D sky, with the third (line-of-sight) direction being frequency dependent. The line-of-sight axis encodes evolution along the light-cone, which can impact the observed 21 cm PS, when compared to predictions from co-evolution cubes
    \citep[e.g.][]{Datta:2012p7679,Datta:2014p4990,LaPlante:2014p7651,Ghara:2015p7650}.  However, this effect is generally only pronounced on much larger scales than considered here (note that our lowest $k$-modes, $|k| = 0.15$ Mpc$^{-1}$, correspond to a modest $\Delta z \sim 0.1$ at EoR redshifts).  We shall nevertheless return to this in future work, extending \cmmc{} to operate directly on the light-cone. 
}
redshifts ($z=6,\,7,\,8,\,9,\,11,\,13,\,15$ and~17). Note that our choice is relatively arbitrary, taken to span the EoH and EoR in both of our mock observations and is not driven by any computational or numerical reasons. 

For the MCMC sampling, we use smaller boxes than used for the mock observations: 300$^{3}$ Mpc$^{3}$ volume on a 200$^3$ grid smoothed down from a high-resolution 1200$^{3}$ grid.  Both the mock observations and the sampled 21 cm PS have the same voxel resolution ($\sim1.5$~Mpc per voxel). These box and cell sizes were selected to ensure per cent level convergence in the PS for a few randomly sampled astrophysical models.

Finally, it is instructive to provide some timing estimates for \cmmc{}. We note that the public version of \cmfst{} is not optimised for our MCMC framework. Therefore, we heavily streamlined the computation for a single core implementation (since the advantage of \cmmc{} is that a realisation of \cmfst{} is performed for each available CPU core). For the $300^{3}$ Mpc$^{3}$ volume, 200$^{3}$ voxel setup above, our streamlined version of \cmfst{} is $\sim20\times$ faster on a single core\footnote{
This boost in computational efficiency arises owing to the inclusion of additional interpolation tables to remove redundant calculations. Additionally, \cmfst{} is reduced to a single executable removing file I/O and limiting memory overheads.
}. For our astrophysical parameter forecasting, we perform $\sim8\times10^4$ \cmfst{} runs. As we are using the parallelised affine invariant ensemble sampler \emcee{}, we can achieve $\sim8\times10^4$ samplings with the following setup: 400 walkers, each performed for 200 iterations\footnote{We explored several combinations of the walker/iterations configuration to confirm our experimental setup was providing converged results for our MCMC.}. Performed using 200 cores (i.e. 100 physical cores + 100 virtual cores) on a shared memory cluster, such a setup takes $\sim8$ days.

\subsection{Parameter recovery for the \textsc{faint galaxies} mock observation} \label{sec:Faint_forecasts}

 \begin{figure*} 
	\begin{center}
		\includegraphics[trim = 0cm 1.5cm 0cm 0.5cm, scale = 0.49]{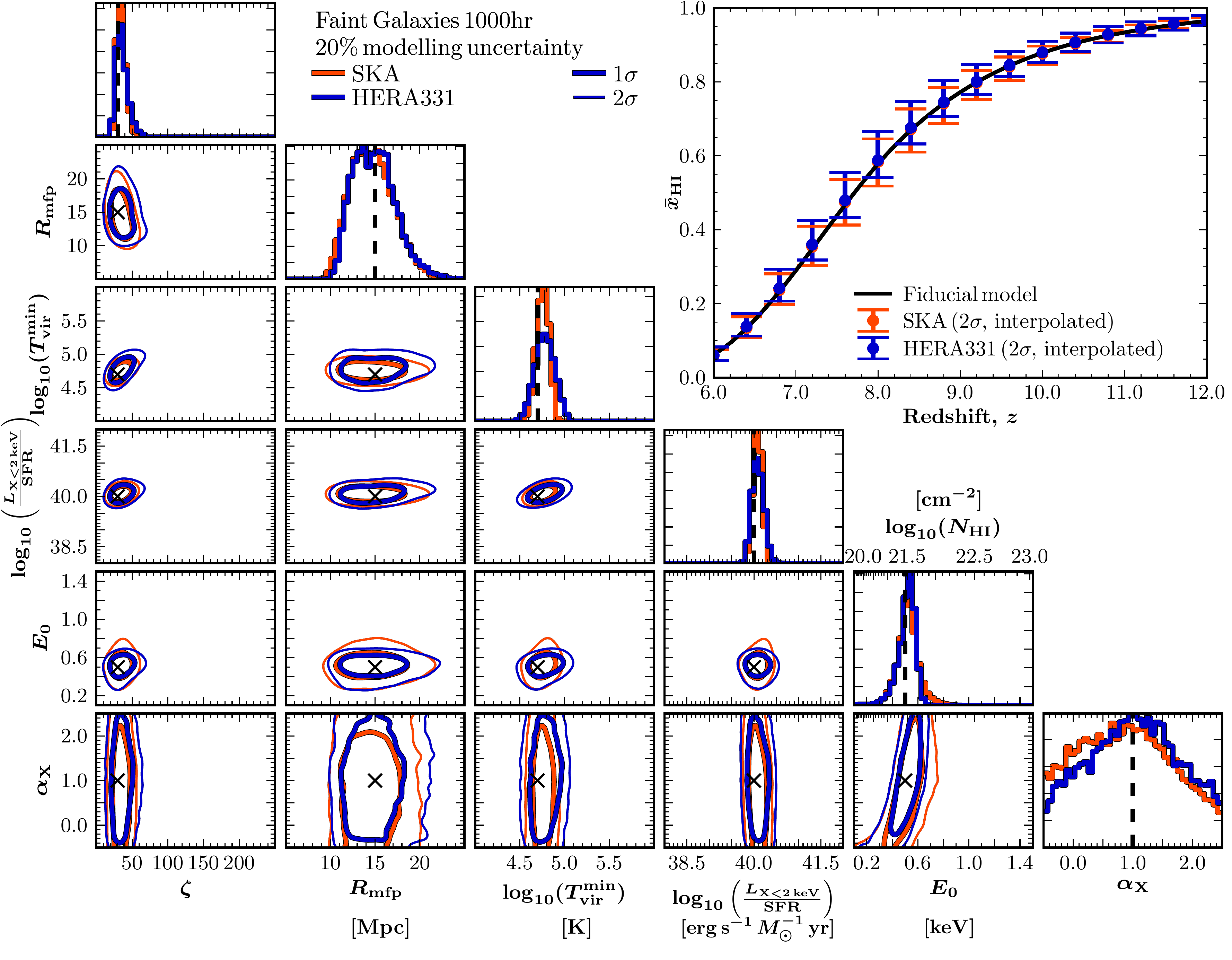}
	\end{center}
\caption[]
{Recovered 1 and 2D marginalised joint posterior distributions for our \textsc{faint galaxies} six parameter astrophysical model for an assumed 1000 hr on sky observation with HERA (blue) and SKA (red). Thick and thin contours correspond to the 68 ($1\sigma$) and 95 ($2\sigma$) per cent 2D marginalised joint likelihood constraints, respectively, and crosses (black vertical dashed lines) denote the input model parameters, defined to be ($\zeta$, $R_{\rm mfp}$, log$_{10}$$(T^{\rm min}_{\rm vir})$, log$_{10}$($L_{\rm X\,<\,2\,keV}/{\rm SFR}$), $E_{0}$, $\alpha_{X}$) = (30, 15, 4.7, 40.0, 0.5, 1.0). Inset: The recovered global evolution of the IGM neutral fraction. The solid black curve corresponds to the fiducial input evolution, whereas the error bars correspond to the $2\sigma$ limits on the recovered IGM neutral fraction. Note, these points are interpolated at $\Delta z = 0.4$ purely for visualisation.}
\label{fig:FaintGalaxies}
\end{figure*}

In Figure~\ref{fig:FaintGalaxies} we present the astrophysical parameter constraints for our \textsc{faint galaxies} model for our assumed 1000hr observation with both HERA (blue) and the SKA (red). Across the diagonals we provide the 1D marginalised probability distribution functions (PDFs) for each of the six model astrophysical parameters. In the lower left half of the figure, we provide the 2D marginalised joint likelihood contours, with crosses denoting the input fiducial values, while the thick and thin contours denote the 68 (1$\sigma$) and 95 (2$\sigma$) percentiles, respectively. In the top right half of the figure, we provide the 2$\sigma$ marginalised constraints on the IGM neutral fraction, $\bar{x}_{\hi{}}$, with respect to the reionisation history of the mock observation (solid black curve). Note that to generate this figure, we interpolate the reionisation history between the marginalised distributions for $\bar{x}_{\hi{}}$ at the eight co-evolution redshifts, sampling at a rate of $\Delta z = 0.4$. In Table~\ref{tab:FaintGalaxies} we provide the median and associated 16th and 84th percentiles for each of our six astrophysical parameters for both HERA and the SKA. 

\begin{table*}
\begin{tabular}{@{}lcccccccccc}
\hline
\textsc{faint galaxies} & & & & Parameter & &  \\
Model/instrument & $\zeta$ & $R_{\rm mfp}$& log$_{10}$$(T^{\rm min}_{\rm vir})$ & log$_{10}$($L_{\rm X\,<\,2\,keV}/{\rm SFR}$) & $E_{0}$ & $\alpha_{X}$ &  \\
 &  & [Mpc] & [K] & [erg s$^{-1}$ $M^{-1}_{\odot}$~yr] & [keV] & \\
\hline
Full X-ray heating \\
\hline
\vspace{0.8mm}
HERA 331 & 34.69$\substack{+9.48 \\ -6.50}$ & 14.82$\substack{+2.47 \\ -2.29}$ & 4.78$\substack{+0.11 \\ -0.11}$ & 40.08$\substack{+0.14 \\ -0.12}$ & 0.52$\substack{+0.08 \\ -0.05}$ & 1.01$\substack{+0.82 \\ -0.85}$ \\
\vspace{0.8mm}
SKA & 33.99$\substack{+7.18 \\ -5.13}$ & 14.56$\substack{+2.58 \\ -2.29}$ & 4.76$\substack{+0.07 \\ -0.07}$ & 40.08$\substack{+0.10 \\ -0.10}$ & 0.52$\substack{+0.07 \\ -0.08}$ & 0.87$\substack{+0.86 \\ -0.85}$ \\
\hline
\multicolumn{2}{@{}l}{Ignoring $T_{\rm S}$ ($T_{\rm S} \gg T_{\cmb}$)} \\
\hline
\vspace{0.8mm}
HERA 331 & 18.98$\substack{+7.81 \\ -2.80}$ & 16.80$\substack{+3.36 \\ -2.67}$ & 4.44$\substack{+0.24 \\ -0.13}$ & - & - & - & \\
\vspace{0.8mm}
SKA & 15.51$\substack{+4.97 \\ -2.20}$ & 16.99$\substack{+3.75 \\ -4.17}$ & 4.29$\substack{+0.17 \\ -0.11}$ & - & - & - & \\
\hline
\end{tabular}
\caption{Summary of the median recovered values (and associated 16th and 84th percentile errors) for the six parameter astrophysical model describing the EoR and EoX, $\zeta$, $R_{\rm mfp}$, log$_{10}$$(T^{\rm min}_{\rm vir})$, log$_{10}$($L_{\rm X\,<\,2\,keV}/{\rm SFR}$), $E_{0}$ and $\alpha_{X}$. We assume a total 1000hr integration time with both the SKA and HERA. Our fiducial mock observation, corresponding to the \textsc{faint galaxies} model, assumes ($\zeta$, $R_{\rm mfp}$, log$_{10}$$(T^{\rm min}_{\rm vir})$, log$_{10}$($L_{\rm X\,<\,2\,keV}/{\rm SFR}$), $E_{0}$, $\alpha_{X}$) = (30, 15, 4.7, 40.0, 0.5, 1.0). We also provide the recovered biased constraints when we ignore the IGM spin temperature fluctuations (i.e. $T_{\rm S} \gg T_{\cmb}$; see Section~\ref{sec:Faint_Bias}).}
\label{tab:FaintGalaxies}
\end{table*}

From the relatively narrow 1D PDFs, it is clear that both HERA and the SKA can simultaneously constrain the EoR and EoH to high accuracy. The only parameter for which we do not achieve strong constraints is the X-ray spectral index, $\alpha_{X}$. However, this is not overly surprising given the relatively small effect it has on the amplitude of the 21 cm PS over the entire allowed parameter range (see Figure~\ref{fig:ParamVary}). Folding in the 20 per cent modelling uncertainty and the instrumental noise, the relative difference in the 21 cm PS amplitude across the full allowed parameter range is roughly consistent at about 2$\sigma$. In terms of the reionisation history both HERA and the SKA recover comparably tight constraints on the IGM neutral fraction. At $1\sigma$, we recover constraints on $\bar{x}_{\hi{}}$ of the order of $\sim5$ per cent.

We recover no strong degeneracies with our astrophysical parameters, with mild degeneracies for $\zeta$-$T^{\rm min}_{\rm vir}$, $L_{\rm X\,<\,2\,keV}/{\rm SFR}$-$T^{\rm min}_{\rm vir}$ and $E_{0}$-$\alpha_{X}$. This is generally consistent with both \citet{EwallWice:2016p6991} and \cite{Kern:2017p8205}, whose mock observation most closely resembles our \textsc{faint galaxies} model. However, these authors find somewhat stronger $f_{\rm X}$-$E_{0}$ and $f_{\rm X}$-$\alpha_{X}$ degeneracies (where $f_{\rm X}$ corresponds to the number of X-ray photons per stellar baryon, and can thus be related to our $L_{\rm X\,<\,2\,keV}/{\rm SFR}$ for a given $\alpha_X$ and $E_0$).  This discrepancy could arise due to (i) the inclusion of the 20 per cent modelling uncertainty which broadens our contours; (ii) the approximations made in those works [assumptions of Gaussian errors in Fisher matrices \citep{EwallWice:2016p6991} or modelling errors in an emulator method \citet{Kern:2017p8205}]; and/or (iii) our choice for the soft-band {\it energy} as a normalisation parameter (instead of the X-ray photon {\it number}), which can provide a more independent basis vector for the EoH evolution\footnote{
In selecting the soft-band X-ray luminosity instead of a harder X-ray band (e.g. 0.5 - 8 keV), we have preferentially minimised the degeneracy between the X-ray luminosity and $\alpha_X$. Additionally, adopting a soft-band X-ray luminosity enables straightforward comparison with numerous observations of nearby galaxies \citep[e.g.][]{Tzanavaris:2008p8147,Mineo:2012p8106,Fragos:2013p6528,Lehmer:2015p7825,Lehmer:2016p7810}. In the near future, we expect observations of the intrinsic soft-band X-ray luminosity escaping the host galaxy to improve with the upcoming Athena telescope \citep{Barcons:2012p8174}, which will provide a soft-band effective area more than an order of magnitude larger that existing experiments (T. Dauser, private communication). This will provide stronger priors on the X-ray SED, even if 21cm observations themselves are less discriminatory.
}.

If we assume the 1D marginalised PDFs can be modelled by a normal distribution, which for the most part is reasonable at the 1$\sigma$ level (i.e. some tails begin to appear at $\sim2\sigma$), we can provide some approximate fractional uncertainties for the astrophysical parameters. For the SKA (HERA), the 1$\sigma$ percentage errors are: $\zeta$ = 18 (24), $R_{\rm mfp}$ = 16 (16), log$_{10}$$(T^{\rm min}_{\rm vir})$ = 1.4 (2.3), log$_{10}$($L_{\rm X\,<\,2\,keV}/{\rm SFR}$) = 0.2 (0.3), $E_{0}$ = 17 (14) and $\alpha_{X}$ = 88 (73).  The uncertainty on the EoR parameters is comparable to what we obtained in \citet{Greig:2015p3675}: $\zeta$ = 17 (22), $R_{\rm mfp}$ = 18 (18), log$_{10}$$(T^{\rm min}_{\rm vir})$ = 2.4 (3.3). Therefore, despite increasing the model complexity by including the EoH, the relative constraints are comparable. 

\citet{EwallWice:2016p6991} quote fractional precisions on their six parameter model, with 1-2 per cent accuracy on the EoR parameters and 6 per cent on their EoH parameters. Their constraints are smaller than ours by about an order of magnitude for the EoR and factor of a few for the EoH. Approximately half of this difference can be attributed to the inclusion of the modelling uncertainty \citep[see e.g.][]{Greig:2015p3675}. The remaining discrepancy can arise from either the fundamental assumptions in their Fisher matrix approach, or their larger number of redshift samples (more than a factor of two). In future, we will modify \cmmc{} to directly work on the observed light cone, removing the necessity of an ad-hoc sampling of co-evolution cubes.

For our \textsc{faint galaxies} model, we find both HERA and the SKA will recover comparable parameter constraints.
This is despite the significantly increased sensitivity achievable with the SKA, resulting in a larger total integrated signal to noise (S/N). As pointed out in \citet{Greig:2015p3675}, the S/N is not a reliable metric for predicting an instrument's ability at parameter constraints, since model constraining power is biased towards large scales.  This highlights the importance of using parameter forecasting as a figure of merit, instead of just the total S/N. We caution however that the SKA performance can be improved if one can better mediate modelling uncertainties.  Indeed, the increased thermal noise sensitivity on small-scales is washed out by our assumed 20 per cent modelling uncertainty. Moreover, the SKA will be superior at tomography; using higher-order likelihood statistics should therefore favour the SKA over HERA.

\subsection{Parameter recovery for the \textsc{bright galaxies} mock observation} \label{sec:Forecast_Bright}

\begin{figure*} 
	\begin{center}
		\includegraphics[trim = 0cm 1.5cm 0cm 0.5cm, scale = 0.49]{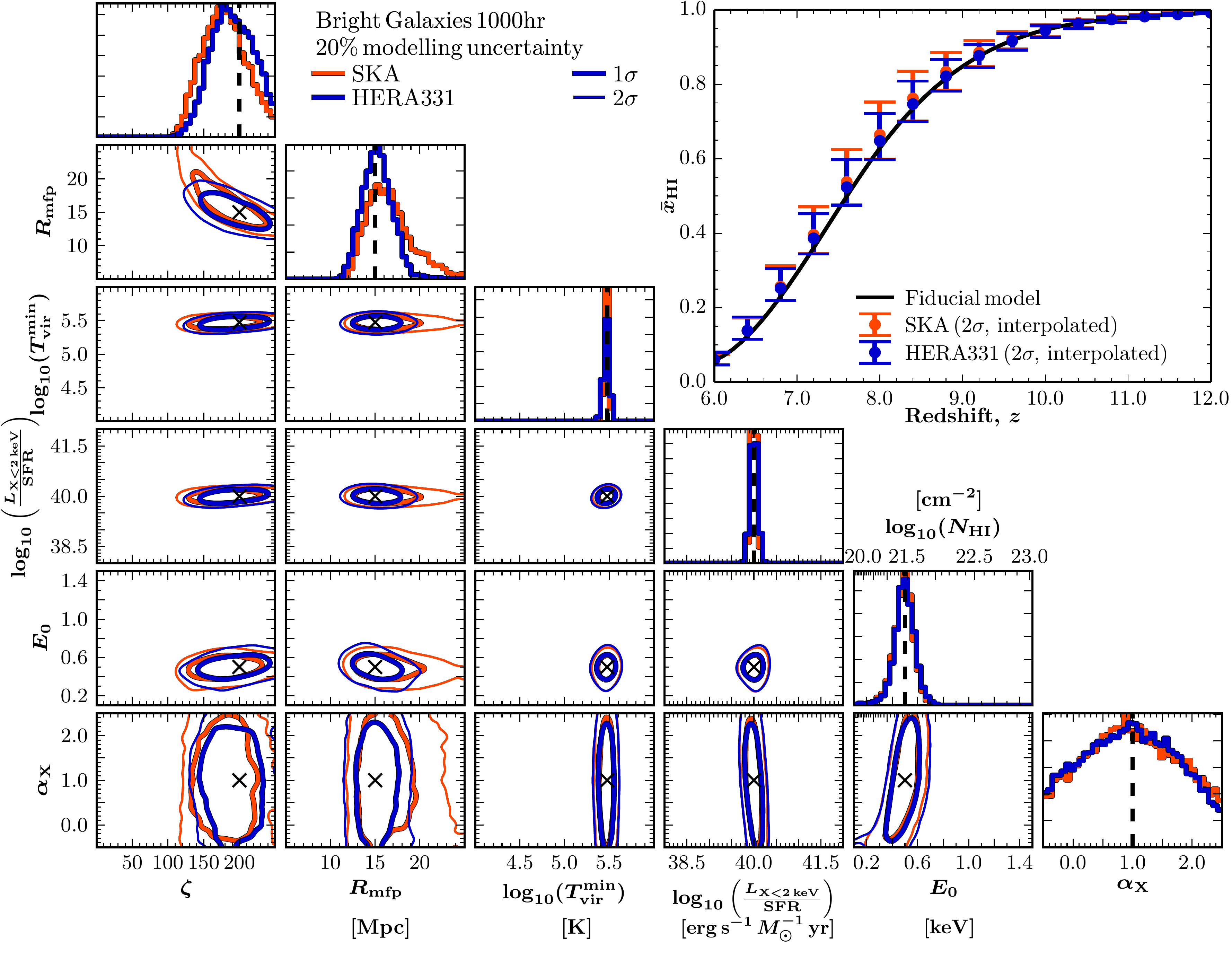}
	\end{center}
\caption[]
{The same as Figure~\ref{fig:FaintGalaxies} except for the \textsc{bright galaxies} model. Crosses (black vertical dashed lines) denote the input model parameters, defined to be ($\zeta$, $R_{\rm mfp}$, log$_{10}$$(T^{\rm min}_{\rm vir})$, log$_{10}$($L_{\rm X\,<\,2\,keV}/{\rm SFR}$), $E_{0}$, $\alpha_{X}$) = (200, 15, 5.48, 40.0, 0.5, 1.0).}
\label{fig:BrightGalaxies}
\end{figure*}

In Figure~\ref{fig:BrightGalaxies} we present our 1 and 2D joint marginalised posterior distributions for each of the six astrophysical parameters for our \textsc{bright galaxies} model assuming a 1000hr observation with HERA (blue) and the SKA (red). Table~\ref{tab:BrightGalaxies} provides the median and associated 16th and 84th percentiles for each of our astrophysical model parameters. As in the previous section, we provide approximate fractional precisions on the model parameters assuming normally distributed marginal likelihoods. For the SKA (HERA), the 1$\sigma$ percent errors are:
$\zeta$ = 17 (15), $R_{\rm mfp}$ = 16 (12), log$_{10}$$(T^{\rm min}_{\rm vir})$ = 0.4 (0.6), 
log$_{10}$($L_{\rm X\,<\,2\,keV}/{\rm SFR}$) = 0.2 (0.2), $E_{0}$ = 16 (17) and $\alpha_{X}$ = 80 (79).

\begin{table*}
\begin{tabular}{@{}lcccccccccc}
\hline
\textsc{bright galaxies} & \multicolumn{6}{c}{Parameter}  \\
Model/instrument & $\zeta$ & $R_{\rm mfp}$& log$_{10}$$(T^{\rm min}_{\rm vir})$ & log$_{10}$($L_{\rm X\,<\,2\,keV}/{\rm SFR}$) & $E_{0}$ & $\alpha_{X}$ &  \\
 &  & [Mpc] & [K] & [erg s$^{-1}$ $M^{-1}_{\odot}$~yr] & [keV] &  \\
\hline
Full X-ray heating \\
\hline
\vspace{0.8mm}
HERA 331 & 188.59$\substack{+32.66 \\ -29.73}$ & 15.21$\substack{+1.75 \\ -1.65}$ & 5.47$\substack{+0.03 \\ -0.03}$ & 40.00$\substack{+0.08 \\ -0.08}$ & 0.50$\substack{+0.08 \\ -0.07}$ & 0.95$\substack{+0.84 \\ -0.86}$ \\
\vspace{0.8mm}
SKA & 177.32$\substack{+33.62 \\ -30.85}$ & 16.33$\substack{+2.91 \\ -2.08}$ & 5.47$\substack{+0.02 \\ -0.02}$ & 40.00$\substack{+0.07 \\ -0.07}$ & 0.50$\substack{+0.07 \\ -0.08}$ & 0.95$\substack{+0.91 \\ -0.82}$ \\
\hline
\multicolumn{2}{@{}l}{Ignoring $T_{\rm S}$ ($T_{\rm S} \gg T_{\cmb}$)} \\
\hline
\vspace{0.8mm}
HERA 331 & 25.50$\substack{+12.42 \\ -4.68}$ & 13.15$\substack{+1.25 \\ -1.19}$ & 4.55$\substack{+0.22 \\ -0.13}$ & - & - & - & \\
\vspace{0.8mm}
SKA & 15.65$\substack{+3.47 \\ -2.22}$ & 11.17$\substack{+2.06 \\ -1.67}$ & 4.18$\substack{+0.13 \\ -0.11}$ & - & - & - & \\
\hline
\end{tabular}
\caption{The same as Table~\ref{tab:FaintGalaxies} except now for the \textsc{bright galaxies} model. Our \textsc{bright galaxies} mock observation assumes ($\zeta$, $R_{\rm mfp}$, log$_{10}$$(T^{\rm min}_{\rm vir})$, log$_{10}$($L_{\rm X\,<\,2\,keV}/{\rm SFR}$), $E_{0}$, $\alpha_{X}$) = (200, 15, 5.48, 40.0, 0.5, 1.0).}
\label{tab:BrightGalaxies}
\end{table*}

The constraints are comparable for most of the astrophysical parameters held fixed across the two models (i.e. $R_{\rm mfp}$, log$_{10}$($L_{\rm X\,<\,2\,keV}/{\rm SFR}$), $E_{0}$, $\alpha_{X}$).
The largest difference is in $R_{\rm mfp}$, which shows tighter constraints relative to the \textsc{faint galaxies} model. At a given stage in the EoR, the \hii{} regions in the \textsc{bright galaxies} model are larger and more isolated, sourced by the brighter, rarer, more-biased sources.  Recombinations play a larger role in limiting the  growth of such \hii{} regions, whose characteristic sizes approach $R_{\rm mfp}$ earlier in reionisation.  Since the signal is more sensitive to $R_{\rm mfp}$, a degeneracy between $\zeta$ and $R_{\rm mfp}$ emerges, with both now able to control the timing of the EoR: one can compensate
for the slowing-down of the EoR due to recombinations (a smaller $R_{\rm mfp}$) by increasing the ionising efficiency, $\zeta$. This degeneracy leads to poorer overall constraints on $\zeta$ when compared to the \textsc{faint galaxies} model.

These improvements in the constraints of $R_{\rm mfp}$ are however only available with HERA. The marginalised PDFs for the SKA exhibit a noticeable tail towards increasing $R_{\rm mfp}$. The source of this tail can simply be attributed to the reduced large-scale sensitivity of the SKA relative to HERA owing to our adopted observing strategies within this work (see Section~\ref{sec:noise} and Figure~\ref{fig:BrightGalaxies_Sensitivity}).

\section{Can we ignore the spin temperature in EoR parameter recovery?} \label{sec:Bias}

The majority of studies that constrain the EoR with the cosmic 21 cm signal assume that the IGM spin temperature is saturated (i.e. $T_{\rm S}\gg T_{\rm CMB}$).   Since the corresponding term in the brightness temperature saturates to unity in this regime (c.f. equation~\ref{eq:21cmTb}), assuming saturation greatly simplifies the computational load.  However, the validity of such an approximation is highly dependent on the poorly-constrained relative efficiencies of ionising and X-ray sources in the first galaxies. Not properly taking into account the IGM spin temperature can bias the recovered EoR parameter constraints\footnote{
In \citet{EwallWice:2016p6991}, these authors found that by incorrectly accounting for the EoH, their recovered uncertainties in the EoR parameters could be biased. They explored this by fitting their EoR parameters by either (i) assuming a fixed EoH model; or (ii) properly fitting and marginalising over the EoH model parameters. The fractional uncertainties from (i) were smaller than (ii) by almost a factor of two. This emphasises the importance of properly accounting for the EoH when performing astrophysical parameter recovery from the 21 cm signal.
}. Here, we quantify the impact of this using our two mock 21 cm observations. The two adopted extrema in $T^{\rm min}_{\rm vir}$ allow the exploration of vastly different levels of overlap between the EoR and EoH, enabling us to estimate the available span in the corresponding bias. Here we use \cmmc{} to recover the fractional precision on the EoR parameters under the simplification of $T_{\rm S}\gg T_{\rm CMB}$, and compare these to the constraints obtained by properly modelling $T_{\rm S}$. 

\subsection{\textsc{Faint galaxies}} \label{sec:Faint_Bias}

\begin{figure*} 
	\begin{center}
		\includegraphics[trim = 0.2cm 1.3cm 0cm 0.7cm, scale = 0.49]{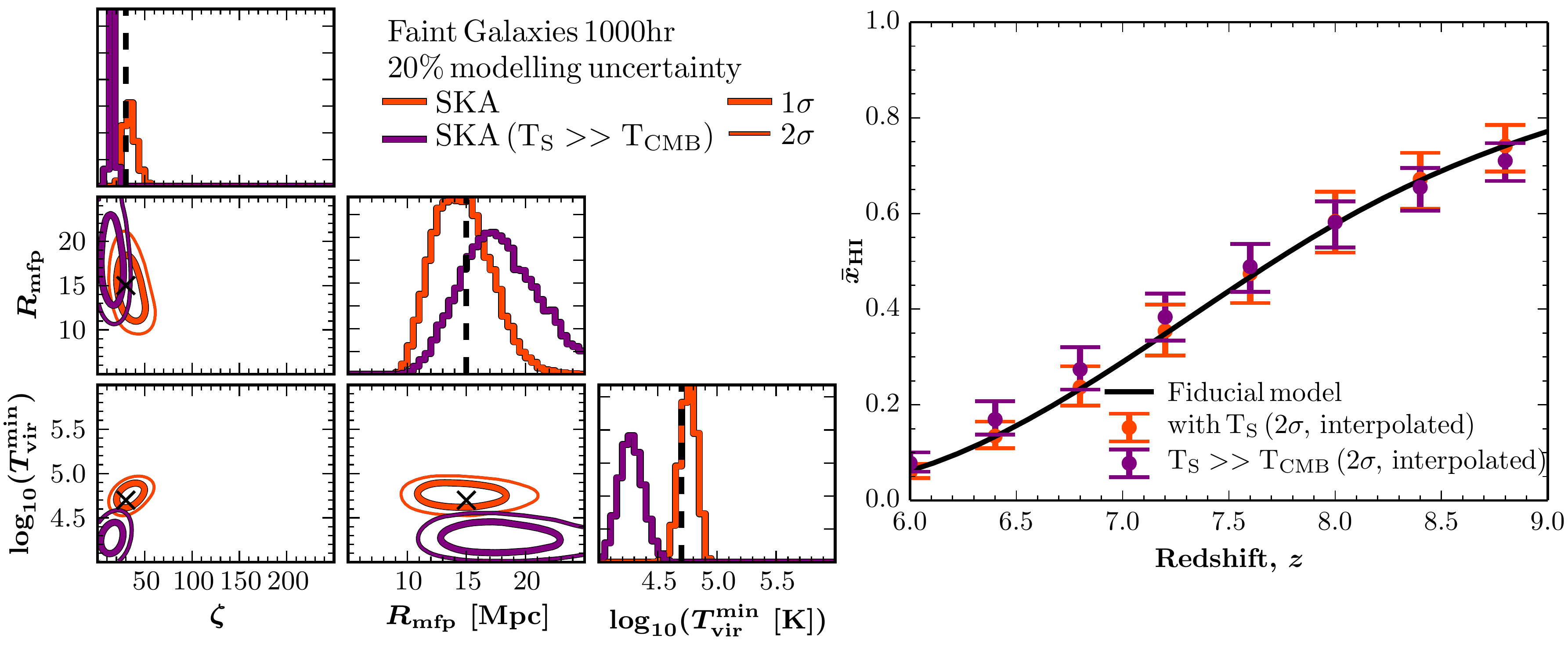}
	\end{center}
\caption[]
        {The impact of the common saturated spin temperature approximation ($T_{\rm S} \gg T_{\rm CMB}$) on EoR parameter inference. We show the recovered 1 and 2D joint marginalised posterior distributions for the \textsc{faint galaxies} model assuming a 1000 hr on sky observation with the SKA. Red curves correspond to our fiducial constraints which include the spin temperature modelling (marginalising over the EoH parameter) accounting for the IGM spin temperature fluctuations (Figure~\ref{fig:FaintGalaxies}), whereas the purple curves are the recovered constraints when ignoring the IGM spin temperature fluctuations (i.e. $T_{\rm S} \gg T_{\cmb}$). Thick and thin contours correspond to the 68 ($1\sigma$) and 95 ($2\sigma$) per cent marginalised joint likelihood contours, respectively, and crosses (black vertical dashed lines) denote the input model parameters, defined to be ($\zeta$, $R_{\rm mfp}$, log$_{10}$$(T^{\rm min}_{\rm vir})$) = (30, 15, 4.7). Inset: The recovered global evolution of the IGM neutral fraction. The solid black curve corresponds to the fiducial input evolution, whereas the error bars correspond to the $2\sigma$ limits on the recovered IGM neutral fraction. Note, all points are interpolated at $\Delta z = 0.4$ purely for visualisation purposes only.
}
\label{fig:FaintGalaxies_Bias}
\end{figure*}

The 21 cm signal for the \textsc{faint galaxies} model transitions from absorption to emission at $z\lesssim12$ (see e.g. the middle panel of Figure~\ref{fig:MockObservations}). At lower redshifts, the IGM spin temperature continues to increase and the $(1-T_{\rm CMB}/T_{S})$ factor approaches unity. Assuming the saturated limit ($T_{\rm S} \gg T_{\rm CMB}$) results in a fractional error in the power spectrum less than 10\% when $(1-T_{\rm CMB}/T_{S})^{2} \gtrsim 0.9$, corresponding to $z\sim7.3$ and an IGM neutral fraction of $\bar{x}_{\hi{}} \sim 0.4$ (cf. the similar \textsc{faint galaxies} model considered in \citealt{Mesinger:2016p6167}). Therefore, the saturated spin temperature approximation is only reasonable during the second half of reionisation in our \textsc{faint galaxies} model.

In Figure~\ref{fig:FaintGalaxies_Bias}, we present the 1 and 2D joint marginalised posterior distributions for the EoR parameters, namely $\zeta$, $R_{\rm mfp}$ and log$_{10}(T^{\rm min}_{\rm vir})$. For this comparison, we only consider the SKA (the results for HERA are nearly identical as shown in Table~\ref{tab:FaintGalaxies}), and run \cmmc{} on the redshifts spanning the EoR: $z = 6,\,7,\,8$ and~9. Red curves correspond to the constraints on the EoR model parameters from fitting the full six parameter astrophysical model (i.e. including the EoH, Figure~\ref{fig:FaintGalaxies}) whereas the purple curves correspond to the constraints obtained with the saturated spin temperature approximation. In the lower half of Table~\ref{tab:FaintGalaxies} we provide the median, 16th and 84th percentiles for the recoverer uncertainties on the EoR parameters.

It is clear from Figure~\ref{fig:FaintGalaxies_Bias} that the saturated spin temperature approximation significantly biases parameter constraints. The saturated temperature approximation prefers models with a smaller virial temperature and ionising efficiency compared to the `true' ones. The marginalised 1D PDFs for $T^{\rm min}_{\rm vir}$ are offset by at least $\sim3\sigma$. For $\zeta$, we recover more modest offsets of $\sim1.5\sigma$, and for $R_{\rm mfp}$ we recover comparable constraints. In all cases, the recovered astrophysical constraints return larger uncertainties, owing to the lower marginalised likelihoods. For reference, the maximum likelihood (ML: $\mathcal{L} = {\rm exp}(-\frac{1}{2}\chi^{2})$) is a factor of four lower than obtained with the full spin temperature modelling in Section~\ref{sec:Faint_forecasts}. 

\begin{figure*} 
	\begin{center}
		\includegraphics[trim = 0cm 0.7cm 0cm 0.5cm, scale = 0.7]{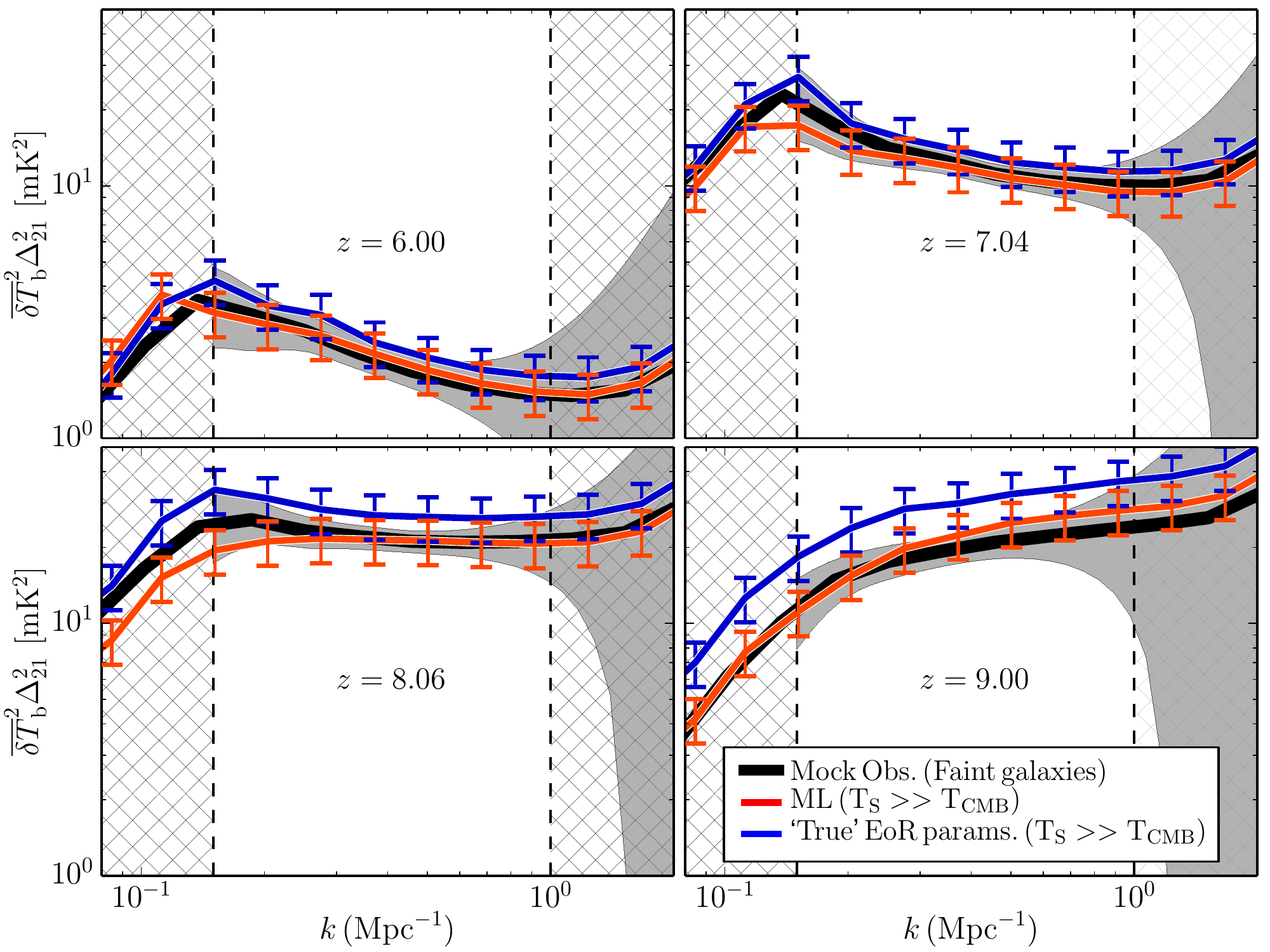}
	\end{center}
\caption[]
        {The 21 cm PS corresponding to the maximum likelihood model assuming a saturated spin temperature ($T_{\rm S} \gg T_{\cmb}$; red curve), compared to the fiducial mock 21 cm PS of the \textsc{faint galaxies} model (black curve). The blue curve corresponds to the ``true'' EoR model parameters, ($\zeta$, $R_{\rm mfp}$, log$_{10}$$(T^{\rm min}_{\rm vir})$) = (30, 15, 4.7), but is computed assuming a saturated spin temperature. The grey shaded region corresponds to the 1$\sigma$ observational uncertainty for an assumed 1000hr observation with the SKA, while the error bars denote our assumed 20 per cent modelling uncertainty on the 21 cm PS. Hatched regions denote $k$-modes outside of our nominal fitting range.}
\label{fig:EoR_Bias_Faint}
\end{figure*}

To understand this bias, in Figure~\ref{fig:EoR_Bias_Faint} we show the 21 cm PS for the mock observation (black curve), the ML 21 cm PS assuming $T_{\rm S} \gg T_{\rm CMB}$ (red curve) and the 21 cm PS adopting the `true' EoR model parameters assuming $T_{\rm S} \gg T_{\rm CMB}$ (blue curve). Error bars denote the 20 per cent modelling uncertainty on the model 21 cm PS. The main impact of ignoring the spin temperature can be seen by comparing the blue and black curves. Since both curves are generated from the same EoR model parameters, any discrepancies arise from the spin temperature. The largest discrepancy, as expected, arises at the highest redshift, where the IGM is still undergoing the final stages of X-ray heating. As discussed in \citet{Mesinger:2016p6167}, during the EoR the cosmic \hi{} patches effectively have the same, uniform IGM spin temperature, resulting in fairly negligible temperature fluctuations for the \textsc{faint galaxies} model \citep[see also][]{Pober:2015p4328}. However, the amplitude of the 21 cm PS is decreased by a factor of $(1 - T_{\rm CMB}/T_{\rm S})^{2}$ (compare the blue and black curves at $z\sim$ 8--9). In order to mimic this $(1 - T_{\rm CMB}/T_{\rm S})^{2}$ decrease in amplitude, a model which assumes $T_{\rm S} \gg T_{\rm CMB}$ will tend to prefer EoR models with intrinsically less power, i.e. those in which the sources are less biased, having a lower $T^{\rm min}_{\rm vir}$. A lower $T^{\rm min}_{\rm vir}$ implies more abundant ionising sources, which must be compensated for by decreasing $\zeta$ to attempt to recover the correct reionisation history. Therefore, the likelihood peaks at smaller virial temperatures and ionising efficiencies (c.f. the red curve, corresponding to $T^{\rm min}_{\rm vir} = 10^{4.29}$ K and $\zeta = 15.5$).  The full evolution of the PS is unable to be completely reproduced, producing the factor of four lower ML compared to the complete model including the EoH\footnote{Note, that the ML from our six parameter model was computed from eight different redshifts, whereas in the saturated limit we only considered four redshifts. As additional redshift data decreases the unnormalised ML value of a model, the relative differences here are even larger.}.

\subsection{\textsc{Bright galaxies}}

\begin{figure*} 
	\begin{center}
		\includegraphics[trim = 0.2cm 1.0cm 0cm 0.5cm, scale = 0.49]{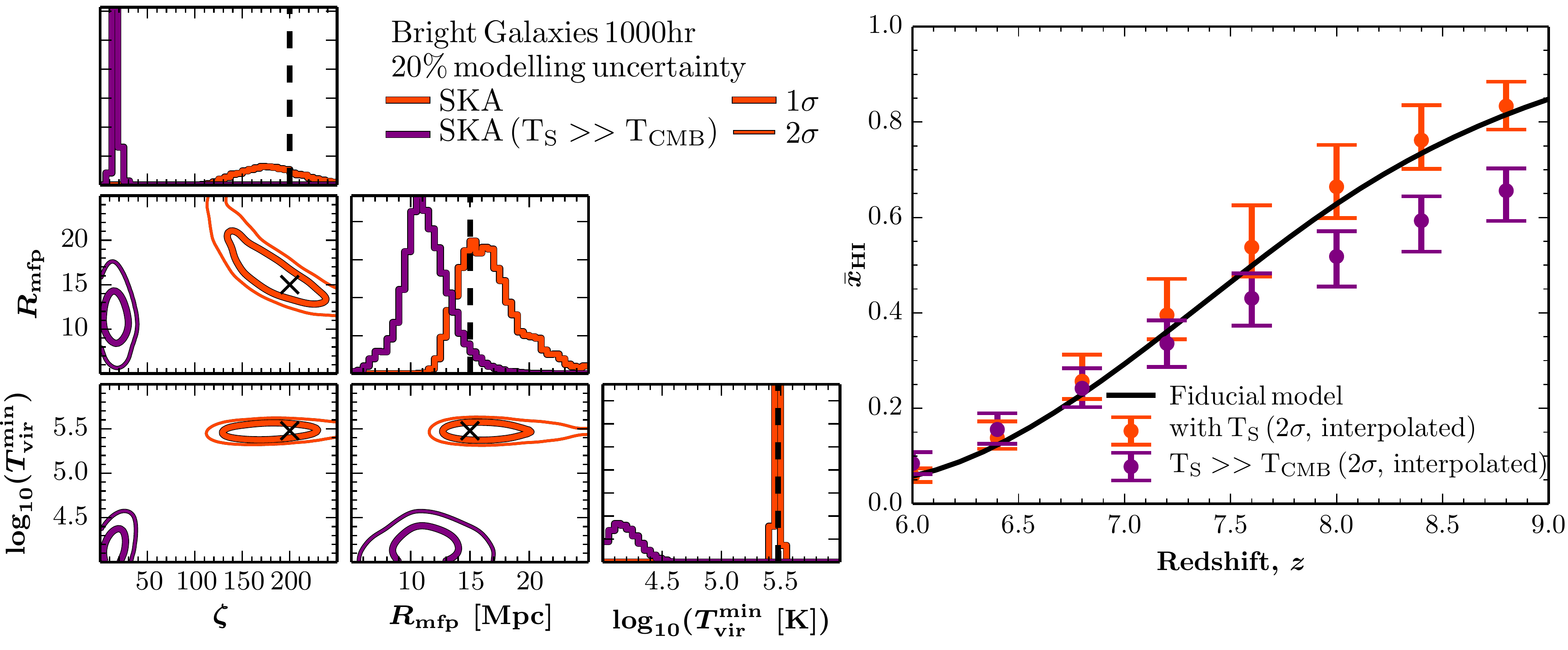}
	\end{center}
\caption[]
{The same as Figure~\ref{fig:FaintGalaxies_Bias}, except for our \textsc{bright galaxies} model. Crosses (black vertical dashed lines) denote the input model parameters, defined to be ($\zeta$, $R_{\rm mfp}$, log$_{10}$$(T^{\rm min}_{\rm vir})$) = (200, 15, 5.48).}
\label{fig:EoR_Bias_Contours_Bright}
\end{figure*}

Already for the \textsc{faint galaxies} model we find significant biases in the recovered EoR parameters under the assumption that the spin temperature is saturated. In the \textsc{bright galaxies} model, the EoH and EoR overlap even more strongly, in effect maximising this bias in parameter recovery. We quantify this in Figure~\ref{fig:EoR_Bias_Contours_Bright}, in which we provide the marginalised distributions, and in the lower half of Table~\ref{tab:BrightGalaxies} where we provide the median, 16th and 84th percentiles for our recovered EoR model parameters.

\begin{figure*} 
	\begin{center}
		\includegraphics[trim = 0cm 0.7cm 0cm 0.5cm, scale = 0.7]{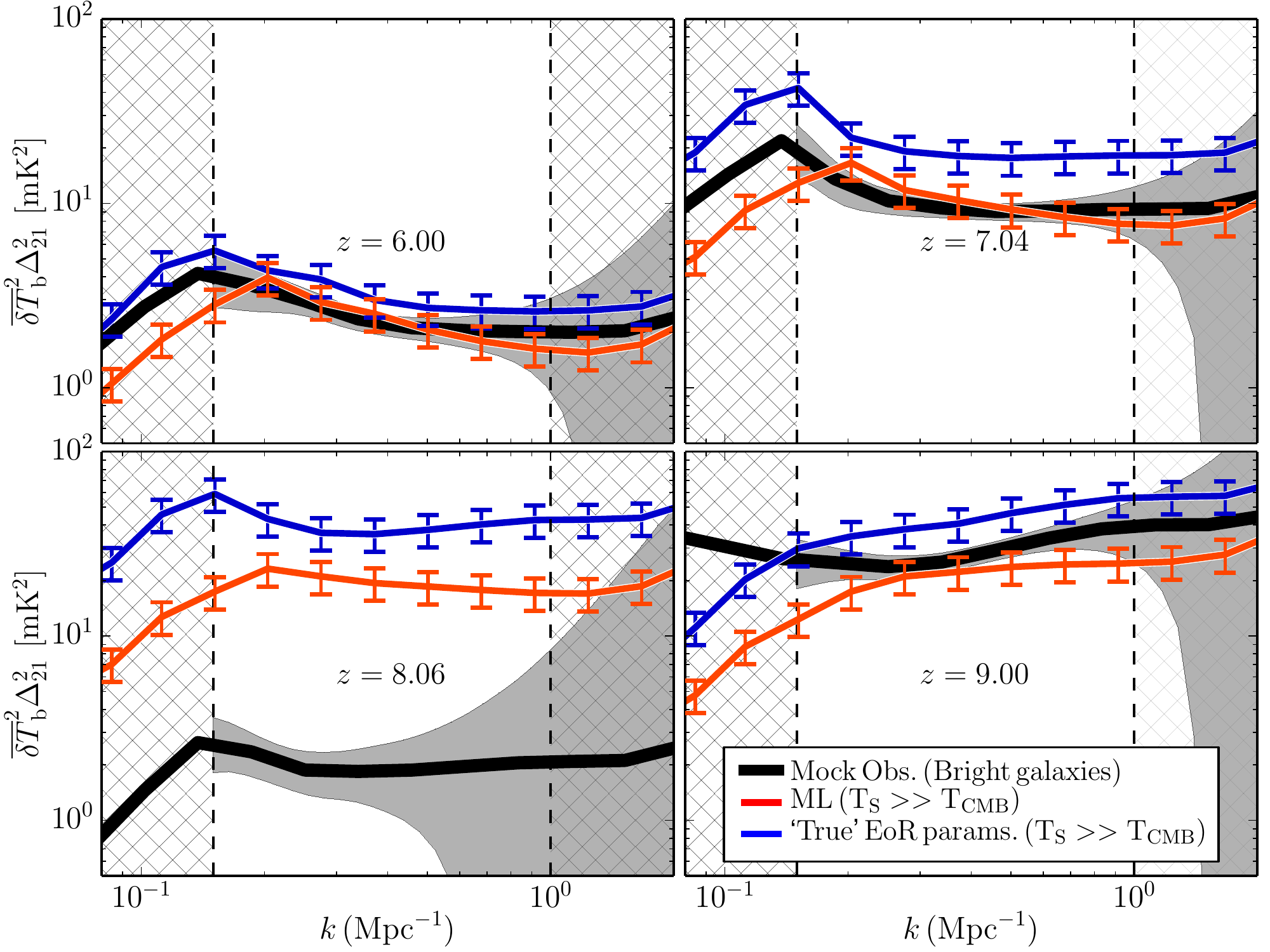}
	\end{center}
\caption[]
{The same as Figure~\ref{fig:EoR_Bias_Faint}, except for our \textsc{bright galaxies} model. The fiducial \textsc{bright galaxies} EoR parameter set corresponds to, ($\zeta$, $R_{\rm mfp}$, log$_{10}$$(T^{\rm min}_{\rm vir})$) = (200, 15, 5.48).}
\label{fig:EoR_Bias_Bright}
\end{figure*}

For both $\zeta$ and $T^{\rm min}_{\rm vir}$ we recover marginalised constraints discrepant at $>10\sigma$. As in the \textsc{faint galaxies} model, $R_{\rm mfp}$ remains consistent to within $1\sigma$, but it instead prefers marginally lower values, $R_{\rm mfp}\sim10$~Mpc. In the inset of Figure~\ref{fig:EoR_Bias_Contours_Bright}, the recovered reionisation history is discrepant at the $>5\sigma$ level, beyond $z\sim7.5$. In Figure~\ref{fig:EoR_Bias_Bright} we present the 21 cm PS from the mock \textsc{bright galaxies} model, the ML estimate and the `true' EoR model parameters assuming the saturated spin temperature limit. It is immediately evident that all the constraining power for the ML arises from the 21 cm PS at $z=6$ and~7. At $z>7$, the saturated limit cannot reproduce the reduced amplitude of the mock 21 cm PS\footnote{For reference, the ML is $\sim10^{28}$ times lower when we assume $T_{\rm S} \gg T_{\rm CMB}$.}.

This is not surprising. As discussed in Section~\ref{sec:Forecast_Bright}, the global averaged 21 cm brightness temperature contrast is still in absorption at $z>8$ (see Figure~\ref{fig:MockObservations}), indicating $T_{\rm S} \lesssim T_{\rm CMB}$ for a significant fraction of the simulation volume. At the same time, the IGM is 35 per cent ionised ($\bar{x}_{\hi{}} = 0.65$) by $z=8$, indicating reionisation is well underway. This significant overlap of the EoR and EoH breaks the fundamental assumption of the saturated spin temperature limit. Furthermore, at $z=8$ the \textsc{bright galaxies} model is transitioning closely to the $T_{\rm S} \equiv T_{\rm CMB}$ limit, producing a precipitous drop in the 21 cm PS amplitude. Relative to the \textsc{bright galaxies} model in the saturated limit (blue curve) this is a factor of $\sim40$ difference in the 21 cm PS amplitude. Under the saturated limit assumption, there is no avenue to mimic such behaviour, resulting in hugely discrepant astrophysical parameter constraints. 

This highlights the importance of properly including the EoH. Incorrectly ignoring the EoH and associated IGM spin temperature fluctuations when interpreting a realistic observation could significantly bias the inferred EoR source model.

\section{Conclusion} \label{sec:conclusion}

Detecting the cosmic 21 cm signal during the EoR and the EoH stands to reveal insights into the formation, growth and evolution of structure in the Universe. However, how do we interpret the underlying astrophysics once we have a detection? To aid this, we developed \cmmc{}, a massively parallel Bayesian MCMC analysis tool, which performs full 3D reionisation simulations (using \cmfst{}) on the fly, for recovering EoR astrophysical parameter constraints. Second generation experiments such as HERA and the SKA, along with global 21 cm experiments, will be able to measure both the EoR and the pre-heating of the IGM by X-rays. Therefore, in order to facilitate simultaneous astrophysical forecasting of the EoR and EoH and consequently aid in the development and construction of the data reduction and signal extraction pipelines, in this work we extend \cmmc{} into the epoch of X-ray heating. 

We demonstrate that both HERA and the SKA will be able to simultaneously constrain the astrophysics of reionisation and the X-ray heating. We consider two models describing the mock observation: \textsc{faint galaxies} and \textsc{bright galaxies}.  These are intended to encompass the physically plausible region of parameter space provided by extrapolating the faint end of the observed UV luminosity function.

Assuming a 1000 hr observation of the 21 cm PS at eight co-evolution redshifts we recover the fractional precision on our six parameter model describing the sources responsible for the EoR and the EoH. These parameters include an ionising source efficiency ($\zeta$), effective photon horizon ($R_{\rm mfp}$), minimum halo mass of ionising sources ($T^{\rm min}_{\rm vir}$), soft-band luminosity ($<2$~keV) of X-ray sources ($L_{\rm X\,<\,2\,keV}/{\rm SFR}$), minimum energy threshold for the attenuation of the X-rays by the host galaxy ISM ($E_{0}$) and the spectral index of the X-ray source SED ($\alpha_{X}$). Additionally assuming a 20 per cent modelling uncertainty in the power spectrum, we recover
the parameters of the mock signals with the following percentage error (1$\sigma$):
\begin{itemize}
\item \textsc{faint galaxies:} $\zeta$ = 18 (24), $R_{\rm mfp}$ = 16 (16), log$_{10}$$(T^{\rm min}_{\rm vir})$ = 1.4 (2.3), 
log$_{10}$($L_{\rm X\,<\,2\,keV}/{\rm SFR}$) = 0.2 (0.3), $E_{0}$ = 17 (14) and $\alpha_{X}$ = 88 (73), respectively for the SKA (HERA).
\item \textsc{bright galaxies:} $\zeta$ = 17 (15), $R_{\rm mfp}$ = 16 (12), log$_{10}$$(T^{\rm min}_{\rm vir})$ = 0.4 (0.6), 
log$_{10}$($L_{\rm X\,<\,2\,keV}/{\rm SFR}$) = 0.2 (0.2), $E_{0}$ = 16 (17) and $\alpha_{X}$ = 80 (79), respectively for the SKA (HERA).
\end{itemize}
Both the SKA and HERA perform equally as well at simultaneously constraining the astrophysics of reionisation and the epoch of X-ray heating due to their comparable sensitivities on the large scales (under our assumptions regarding the survey strategy) which most strongly discriminate between the astrophysical models.

With our expanded framework, we also quantify the impact of the common assumption of a saturated spin temperature, $T_{\rm S}\gg T_{\rm CMB}$, during the EoR.
Our \textsc{faint galaxies} model has a relatively distinct EoH and EoR, for which we would typically expect $T_{\rm S}\gg T_{\rm CMB}$ to be a reasonable approximation during the bulk of reionisation. Nevertheless, even this modest overlap of epochs leads to biases in $\zeta$ and $T^{\rm min}_{\rm vir}$ of up to $\sim3\sigma$, and a reduction in the ML by a factor of $>4$. The \textsc{bright galaxies} model on the other hand represents the extreme case of EoR and EoH overlap. For this model, the recovered constraints are discrepant at $>10\sigma$, with the ML under in the saturated limit being a factor of  $\sim10^{28}$ lower than in the full model.
Therefore, adopting the saturated spin temperature approximation can significantly bias inferences on EoR parameters.
  
\section*{Acknowledgements}
We thank Adrian Liu and Nicholas Kern for comments on a draft version of this work.  This work was supported by the European Research Council (ERC) under the European Union's Horizon 2020 research and innovation programme (grant agreement No 638809 -- AIDA -- PI: Mesinger).

\bibliography{21CMMC_wTs}

\begin{thebibliography}{138}
\expandafter\ifx\csname natexlab\endcsname\relax\def\natexlab#1{#1}\fi

\bibitem[{Abel {et~al.}(2002)Abel, Bryan, \& Norman}]{Abel:2002p2149}
Abel T., Bryan G.~L., Norman M.~L., 2002, Science, 295, 93

\bibitem[{Akeret {et~al.}(2013)Akeret, Seehars, Amara, Refregier, \&
  Csillaghy}]{Akeret:2012p842}
Akeret J., Seehars S., Amara A., Refregier A., Csillaghy A., 2013, Astron.
  Comput., 2, 27

\bibitem[{Ali {et~al.}(2015)}]{Ali:2015p4327}
Ali Z.~S., {et~al.}, 2015, ApJ, 809, 61

\bibitem[{Alvarez \& Abel(2012)}]{Alvarez:2012p1930}
Alvarez M.~A., Abel T., 2012, ApJ, 747, 126

\bibitem[{Baek {et~al.}(2010)Baek, Semelin, Matteo, Revaz, \&
  Combes}]{Baek:2010p6357}
Baek S., Semelin B., Matteo P.~D., Revaz Y., Combes F., 2010, A\&A, 523, A4

\bibitem[{Barcons {et~al.}(2012)}]{Barcons:2012p8174}
Barcons X., {et~al.}, 2012, preprint (arXiv:1207.2745)

\bibitem[{Barkana \& Loeb(2001)}]{Barkana:2001p1634}
Barkana R., Loeb A., 2001, Phys. Rep., 349, 125

\bibitem[{Barkana \& Loeb(2005)}]{Barkana:2005p1934}
---, 2005, ApJ, 626, 1

\bibitem[{Barkana \& Loeb(2007)}]{Barkana:2007p2929}
---, 2007, Rep. Prog. Phys., 70, 627

\bibitem[{Barone-Nugent {et~al.}(2014)}]{BaroneNugent:2014p4324}
Barone-Nugent R.~L., {et~al.}, 2014, ApJ, 793, 17

\bibitem[{Beardsley {et~al.}(2014)Beardsley, Morales, Lidz, Malloy, \&
  Sutter}]{Beardsley:2014p1529}
Beardsley A., Morales M., Lidz A., Malloy M., Sutter P., 2014, preprint
  (arXiv:1410.5427)

\bibitem[{Behroozi \& Silk(2015)}]{Behroozi:2015p1}
Behroozi P.~S., Silk J., 2015, ApJ, 799, 32

\bibitem[{Bernardi {et~al.}(2016)}]{Bernardi:2016p6834}
Bernardi G., {et~al.}, 2016, MNRAS, 461, 2847

\bibitem[{Bolton \& Haehnelt(2007)}]{Bolton:2007p3273}
Bolton J.~S., Haehnelt M.~G., 2007, MNRAS, 382, 325

\bibitem[{Bond {et~al.}(1991)Bond, Cole, Efstathiou, \& Kaiser}]{Bond:1991p111}
Bond J.~R., Cole S., Efstathiou G., Kaiser N., 1991, ApJ, 379, 440

\bibitem[{Bouwens {et~al.}(2015)}]{Bouwens:2015p7832}
Bouwens R.~J., {et~al.}, 2015, ApJ, 803, 34

\bibitem[{Bowman \& Rogers(2010)}]{Bowman:2010p6724}
Bowman J.~D., Rogers A. E.~E., 2010, Nature, 468, 796

\bibitem[{Bromm {et~al.}(2002)Bromm, Coppi, \& Larson}]{Bromm:2002p2153}
Bromm V., Coppi P.~S., Larson R.~B., 2002, ApJ, 564, 23

\bibitem[{Burns {et~al.}(2012)}]{Burns:2012p6941}
Burns J.~O., {et~al.}, 2012, AdSpR, 49, 433

\bibitem[{Cirelli {et~al.}(2009)Cirelli, Iocco, \& Panci}]{Cirelli:2009p6569}
Cirelli M., Iocco F., Panci P., 2009, JCAP, 10, 009

\bibitem[{Das {et~al.}(2017)Das, Mesinger, Pallottini, Ferrara, \&
  Wise}]{Das:2017p7170}
Das A., Mesinger A., Pallottini A., Ferrara A., Wise J.~H., 2017, preprint
  (arXiv:1702.00409)

\bibitem[{Datta {et~al.}(2010)Datta, Bowman, \& Carilli}]{Datta:2010p2792}
Datta A., Bowman J.~D., Carilli C.~L., 2010, ApJ, 724, 526

\bibitem[{Datta {et~al.}(2014)Datta, Jensen, Majumdar, Mellema, Iliev, Mao,
  Shapiro, \& Ahn}]{Datta:2014p4990}
Datta K.~K., Jensen H., Majumdar S., Mellema G., Iliev I.~T., Mao Y., Shapiro
  P.~R., Ahn K., 2014, MNRAS, 442, 1491

\bibitem[{Datta {et~al.}(2012)Datta, Mellema, Mao, Iliev, Shapiro, \&
  Ahn}]{Datta:2012p7679}
Datta K.~K., Mellema G., Mao Y., Iliev I.~T., Shapiro P.~R., Ahn K., 2012,
  MNRAS, 424, 1877

\bibitem[{DeBoer {et~al.}(2017)}]{DeBoer:2017p6740}
DeBoer D.~R., {et~al.}, 2017, PASP, 129, 045001

\bibitem[{Dijkstra {et~al.}(2012)Dijkstra, Gilfanov, Loeb, \&
  Sunyaev}]{Dijkstra:2012p7165}
Dijkstra M., Gilfanov M., Loeb A., Sunyaev R., 2012, MNRAS, 421, 213

\bibitem[{Evoli {et~al.}(2014)Evoli, Mesinger, \& Ferrara}]{Evoli:2014p6707}
Evoli C., Mesinger A., Ferrara A., 2014, JCAP, 11, 024

\bibitem[{Ewall-Wice {et~al.}(2016)Ewall-Wice, Hewitt, Mesinger, Dillon, Liu,
  \& Pober}]{EwallWice:2016p6991}
Ewall-Wice A., Hewitt J., Mesinger A., Dillon J.~S., Liu A., Pober J., 2016,
  MNRAS, 458, 2710

\bibitem[{Fialkov {et~al.}(2014)Fialkov, Barkana, \&
  Visbal}]{Fialkov:2014p3720}
Fialkov A., Barkana R., Visbal E., 2014, Nature, 506, 197

\bibitem[{Fialkov {et~al.}(2013)Fialkov, Barkana, Visbal, Tseliakhovich, \&
  Hirata}]{Fialkov:2013p2903}
Fialkov A., Barkana R., Visbal E., Tseliakhovich D., Hirata C.~M., 2013, MNRAS,
  432, 2909

\bibitem[{Fialkov {et~al.}(2017)Fialkov, Cohen, Barkana, \&
  Silk}]{Fialkov:2017p6998}
Fialkov A., Cohen A., Barkana R., Silk J., 2017, MNRAS, 464, 3498

\bibitem[{Field(1958)}]{Field:1958p1}
Field G.~B., 1958, Proc. Inst. Radio Eng., 46, 240

\bibitem[{Foreman-Mackey {et~al.}(2013)Foreman-Mackey, Hogg, Lang, \&
  Goodman}]{ForemanMackey:2013p823}
Foreman-Mackey D., Hogg D.~W., Lang D., Goodman J., 2013, PASP, 125, 306

\bibitem[{Fragos {et~al.}(2013{\natexlab{a}})Fragos, Lehmer, Naoz, Zezas, \&
  Basu-Zych}]{Fragos:2013p6528}
Fragos T., Lehmer B.~D., Naoz S., Zezas A., Basu-Zych A., 2013{\natexlab{a}},
  ApJL, 776, L31

\bibitem[{Fragos {et~al.}(2013{\natexlab{b}})}]{Fragos:2013p6529}
Fragos T., {et~al.}, 2013{\natexlab{b}}, ApJ, 764, 41

\bibitem[{Furlanetto(2006)}]{Furlanetto:2006p3782}
Furlanetto S.~R., 2006, MNRAS, 371, 867

\bibitem[{Furlanetto \& Mesinger(2009)}]{Furlanetto:2009p7171}
Furlanetto S.~R., Mesinger A., 2009, MNRAS, 394, 1667

\bibitem[{Furlanetto \& Oh(2005)}]{Furlanetto:2005p4326}
Furlanetto S.~R., Oh S.~P., 2005, MNRAS, 363, 1031

\bibitem[{Furlanetto {et~al.}(2006)Furlanetto, Oh, \&
  Briggs}]{Furlanetto:2006p209}
Furlanetto S.~R., Oh S.~P., Briggs F.~H., 2006, Phys. Rep., 433, 181

\bibitem[{Furlanetto {et~al.}(2004)Furlanetto, Zaldarriaga, \&
  Hernquist}]{Furlanetto:2004p123}
Furlanetto S.~R., Zaldarriaga M., Hernquist L., 2004, ApJ, 613, 1

\bibitem[{Ghara {et~al.}(2016)Ghara, Choudhury, \& Datta}]{Ghara:2016p6706}
Ghara R., Choudhury T.~R., Datta K.~K., 2016, MNRAS, 460, 827

\bibitem[{Ghara {et~al.}(2015)Ghara, Datta, \& Choudhury}]{Ghara:2015p7650}
Ghara R., Datta K.~K., Choudhury T.~R., 2015, MNRAS, 453, 3143

\bibitem[{Gnedin \& Ostriker(1997)}]{Gnedin:1997p4494}
Gnedin N.~Y., Ostriker J.~P., 1997, ApJ, 486, 581

\bibitem[{Gnedin \& Shaver(2004)}]{Gnedin:2004p4481}
Gnedin N.~Y., Shaver P.~A., 2004, ApJ, 608, 611

\bibitem[{Goodman \& Weare(2010)}]{Goodman:2010p843}
Goodman J., Weare J., 2010, Commun. Appl. Math. Comput. Sci., 5, 1

\bibitem[{Greenhill \& Bernardi(2012)}]{Greenhill:2012p6829}
Greenhill L.~J., Bernardi G., 2012, preprint (arXiv:1201.1700)

\bibitem[{Greig \& Mesinger(2015)}]{Greig:2015p3675}
Greig B., Mesinger A., 2015, MNRAS, 449, 4246

\bibitem[{Greig {et~al.}(2015)Greig, Mesinger, \& Koopmans}]{Greig:2015p6161}
Greig B., Mesinger A., Koopmans L. V.~E., 2015, preprint (arXiv:1509.03312)

\bibitem[{Haiman {et~al.}(2000)Haiman, Abel, \& Rees}]{Haiman:2000p2155}
Haiman Z., Abel T., Rees M.~J., 2000, ApJ, 534, 11

\bibitem[{Haiman \& Bryan(2006)}]{Haiman:2006p2169}
Haiman Z., Bryan G.~L., 2006, ApJ, 650, 7

\bibitem[{Haiman {et~al.}(1996)Haiman, Thoul, \& Loeb}]{Haiman:1996p2144}
Haiman Z., Thoul A.~A., Loeb A., 1996, ApJ, 464, 523

\bibitem[{Hazelton {et~al.}(2013)Hazelton, Morales, \&
  Sullivan}]{Hazelton:2013p1481}
Hazelton B.~J., Morales M.~F., Sullivan I.~S., 2013, ApJ, 770, 156

\bibitem[{Holzbauer \& Furlanetto(2012)}]{Holzbauer:2012p2890}
Holzbauer L.~N., Furlanetto S.~R., 2012, MNRAS, 419, 718

\bibitem[{Kern {et~al.}(2017)Kern, Liu, Parsons, Mesinger, \&
  Greig}]{Kern:2017p8205}
Kern N.~S., Liu A., Parsons A.~R., Mesinger A., Greig B., 2017, preprint
  (arXiv:1705.04688)

\bibitem[{Kimm {et~al.}(2017)Kimm, Katz, Haehnelt, Rosdahl, Devriendt, \&
  Slyz}]{Kimm:2017p7875}
Kimm T., Katz H., Haehnelt M., Rosdahl J., Devriendt J., Slyz A., 2017, MNRAS,
  466, 4826

\bibitem[{Kuhlen \& Faucher-Gigu{\`e}re(2012)}]{Kuhlen:2012p1506}
Kuhlen M., Faucher-Gigu{\`e}re C.-A., 2012, MNRAS, 423, 862

\bibitem[{{La Plante} {et~al.}(2014){La Plante}, Battaglia, Natarajan,
  Peterson, Trac, Cen, \& Loeb}]{LaPlante:2014p7651}
{La Plante} P., Battaglia N., Natarajan A., Peterson J.~B., Trac H., Cen R.,
  Loeb A., 2014, ApJ, 789, 31

\bibitem[{Lacey \& Cole(1993)}]{Lacey:1993p115}
Lacey C., Cole S., 1993, MNRAS, 262, 627

\bibitem[{Lehmer {et~al.}(2012)}]{Lehmer:2012p7191}
Lehmer B.~D., {et~al.}, 2012, ApJ, 752, 46

\bibitem[{Lehmer {et~al.}(2013)}]{Lehmer:2013p7818}
---, 2013, ApJ, 771, 134

\bibitem[{Lehmer {et~al.}(2015)}]{Lehmer:2015p7825}
---, 2015, ApJ, 806, 126

\bibitem[{Lehmer {et~al.}(2016)}]{Lehmer:2016p7810}
---, 2016, ApJ, 825, 7

\bibitem[{Leite {et~al.}(2017)Leite, Evoli, D'Angelo, Ciardi, Sigl, \&
  Ferrara}]{Leite:2017p09337}
Leite N., Evoli C., D'Angelo M., Ciardi B., Sigl G., Ferrara A., 2017, preprint
  (arXiv:1703.09337)

\bibitem[{Lidz {et~al.}(2007)Lidz, Zahn, McQuinn, Zaldarriaga, Dutta, \&
  Hernquist}]{Lidz:2007p1929}
Lidz A., Zahn O., McQuinn M., Zaldarriaga M., Dutta S., Hernquist L., 2007,
  ApJ, 659, 865

\bibitem[{Lidz {et~al.}(2008)Lidz, Zahn, McQuinn, Zaldarriaga, \&
  Hernquist}]{Lidz:2008p1744}
Lidz A., Zahn O., McQuinn M., Zaldarriaga M., Hernquist L., 2008, ApJ, 680, 962

\bibitem[{Liu {et~al.}(2014{\natexlab{a}})Liu, Parsons, \&
  Trott}]{Liu:2014p3465}
Liu A., Parsons A.~R., Trott C.~M., 2014{\natexlab{a}}, Phys. Rev. D, 90,
  023018

\bibitem[{Liu {et~al.}(2014{\natexlab{b}})Liu, Parsons, \&
  Trott}]{Liu:2014p3466}
---, 2014{\natexlab{b}}, Phys. Rev. D, 90, 023019

\bibitem[{Loeb \& Furlanetto(2013)}]{Loeb:2013p2936}
Loeb A., Furlanetto S.~R., 2013, {The First Galaxies in the Universe. Princeton
  Univ. Press, Princeton, NJ}

\bibitem[{Lopez-Honorez {et~al.}(2016)Lopez-Honorez, Mena, Molin{\'e},
  Palomares-Ruiz, \& Vincent}]{LopezHonorez:2016p6709}
Lopez-Honorez L., Mena O., Molin{\'e} {\'A}., Palomares-Ruiz S., Vincent A.~C.,
  2016, JCAP, 08, 004

\bibitem[{Madau {et~al.}(1997)Madau, Meiksin, \& Rees}]{Madau:1997p4479}
Madau P., Meiksin A., Rees M.~J., 1997, ApJ, 475, 429

\bibitem[{Madau {et~al.}(2004)Madau, Rees, Volonteri, Haardt, \&
  Oh}]{Madau:2004p6564}
Madau P., Rees M.~J., Volonteri M., Haardt F., Oh S.~P., 2004, ApJ, 604, 484

\bibitem[{McQuinn(2012)}]{McQuinn:2012p3773}
McQuinn M., 2012, MNRAS, 426, 1349

\bibitem[{McQuinn {et~al.}(2007)McQuinn, Lidz, Zahn, Dutta, Hernquist, \&
  Zaldarriaga}]{McQuinn:2007p1665}
McQuinn M., Lidz A., Zahn O., Dutta S., Hernquist L., Zaldarriaga M., 2007,
  MNRAS, 377, 1043

\bibitem[{McQuinn {et~al.}(2011)McQuinn, Oh, \&
  Faucher-Gigu{\`e}re}]{McQuinn:2011p3293}
McQuinn M., Oh S.~P., Faucher-Gigu{\`e}re C.-A., 2011, ApJ, 743, 82

\bibitem[{McQuinn \& O'Leary(2012)}]{McQuinn:2012p3776}
McQuinn M., O'Leary R.~M., 2012, ApJ, 760, 3

\bibitem[{McQuinn {et~al.}(2006)McQuinn, Zahn, Zaldarriaga, Hernquist, \&
  Furlanetto}]{McQuinn:2006p109}
McQuinn M., Zahn O., Zaldarriaga M., Hernquist L., Furlanetto S.~R., 2006, ApJ,
  653, 815

\bibitem[{Mellema {et~al.}(2013)}]{Mellema:2013p2975}
Mellema G., {et~al.}, 2013, Exp. Astron., 36, 235

\bibitem[{Mesinger {et~al.}(2006)Mesinger, Bryan, \&
  Haiman}]{Mesinger:2006p2171}
Mesinger A., Bryan G.~L., Haiman Z., 2006, ApJ, 648, 835

\bibitem[{Mesinger {et~al.}(2013)Mesinger, Ferrara, \&
  Spiegel}]{Mesinger:2013p1835}
Mesinger A., Ferrara A., Spiegel D.~S., 2013, MNRAS, 431, 621

\bibitem[{Mesinger \& Furlanetto(2007)}]{Mesinger:2007p122}
Mesinger A., Furlanetto S., 2007, ApJ, 669, 663

\bibitem[{Mesinger {et~al.}(2011)Mesinger, Furlanetto, \&
  Cen}]{Mesinger:2011p1123}
Mesinger A., Furlanetto S., Cen R., 2011, MNRAS, 411, 955

\bibitem[{Mesinger {et~al.}(2016)Mesinger, Greig, \&
  Sobacchi}]{Mesinger:2016p6167}
Mesinger A., Greig B., Sobacchi E., 2016, MNRAS, 459, 2342

\bibitem[{Mesinger {et~al.}(2012)Mesinger, McQuinn, \&
  Spergel}]{Mesinger:2012p1131}
Mesinger A., McQuinn M., Spergel D.~N., 2012, MNRAS, 422, 1403

\bibitem[{Mineo {et~al.}(2012{\natexlab{a}})Mineo, Gilfanov, \&
  Sunyaev}]{Mineo:2012p6282}
Mineo S., Gilfanov M., Sunyaev R., 2012{\natexlab{a}}, MNRAS, 419, 2095

\bibitem[{Mineo {et~al.}(2012{\natexlab{b}})Mineo, Gilfanov, \&
  Sunyaev}]{Mineo:2012p8106}
---, 2012{\natexlab{b}}, MNRAS, 426, 1870

\bibitem[{Mirabel {et~al.}(2011)Mirabel, Dijkstra, Laurent, Loeb, \&
  Pritchard}]{Mirabel:2011p6518}
Mirabel I.~F., Dijkstra M., Laurent P., Loeb A., Pritchard J.~R., 2011, A\&A,
  528, A149

\bibitem[{Morales(2005)}]{Morales:2005p1474}
Morales M.~F., 2005, ApJ, 619, 678

\bibitem[{Morales {et~al.}(2012)Morales, Hazelton, Sullivan, \&
  Beardsley}]{Morales:2012p2828}
Morales M.~F., Hazelton B., Sullivan I., Beardsley A., 2012, ApJ, 752, 137

\bibitem[{Morales \& Wyithe(2010)}]{Morales:2010p1274}
Morales M.~F., Wyithe J. S.~B., 2010, ARA\&A, 48, 127

\bibitem[{Oh(2001)}]{Oh:2001p6617}
Oh S.~P., 2001, ApJ, 553, 499

\bibitem[{Paardekooper {et~al.}(2015)Paardekooper, Khochfar, \& {Dalla
  Vecchia}}]{Paardekooper:2015p1}
Paardekooper J.-P., Khochfar S., {Dalla Vecchia} C., 2015, MNRAS, 451, 2544

\bibitem[{Pacucci {et~al.}(2014)Pacucci, Mesinger, Mineo, \&
  Ferrara}]{Pacucci:2014p4323}
Pacucci F., Mesinger A., Mineo S., Ferrara A., 2014, MNRAS, 443, 678

\bibitem[{Parsons {et~al.}(2012{\natexlab{a}})Parsons, Pober, McQuinn, Jacobs,
  \& Aguirre}]{Parsons:2012p95}
Parsons A., Pober J., McQuinn M., Jacobs D., Aguirre J., 2012{\natexlab{a}},
  ApJ, 753, 81

\bibitem[{Parsons {et~al.}(2012{\natexlab{b}})Parsons, Pober, Aguirre, Carilli,
  Jacobs, \& Moore}]{Parsons:2012p2833}
Parsons A.~R., Pober J.~C., Aguirre J.~E., Carilli C.~L., Jacobs D.~C., Moore
  D.~F., 2012{\natexlab{b}}, ApJ, 756, 165

\bibitem[{Parsons {et~al.}(2010)}]{Parsons:2010p3000}
Parsons A.~R., {et~al.}, 2010, AJ, 139, 1468

\bibitem[{Parsons {et~al.}(2014)}]{Parsons:2014p781}
---, 2014, ApJ, 788, 106

\bibitem[{Patra {et~al.}(2015)Patra, Subrahmanyan, Sethi, Shankar, \&
  Raghunathan}]{Patra:2015p6814}
Patra N., Subrahmanyan R., Sethi S., Shankar N.~U., Raghunathan A., 2015, ApJ,
  801, 138

\bibitem[{{Planck Collaboration XIII}(2016)}]{PlanckCollaboration:2016p7780}
{Planck Collaboration XIII}, 2016, A\&A, 594, A13

\bibitem[{{Planck Collaboration XLVII}(2016)}]{PlanckCollaboration:2016p7627}
{Planck Collaboration XLVII}, 2016, A\&A, 596, A108

\bibitem[{Pober {et~al.}(2013)}]{Pober:2013p41}
Pober J.~C., {et~al.}, 2013, AJ, 145, 65

\bibitem[{Pober {et~al.}(2014)}]{Pober:2014p35}
---, 2014, ApJ, 782, 66

\bibitem[{Pober {et~al.}(2015)}]{Pober:2015p4328}
---, 2015, ApJ, 809, 62

\bibitem[{Pober {et~al.}(2016)}]{Pober:2016p7301}
---, 2016, ApJ, 819, 8

\bibitem[{Power {et~al.}(2013)Power, James, Combet, \& Wynn}]{Power:2013p6533}
Power C., James G., Combet C., Wynn G., 2013, ApJ, 764, 76

\bibitem[{Power {et~al.}(2009)Power, Wynn, Combet, \&
  Wilkinson}]{Power:2009p6560}
Power C., Wynn G.~A., Combet C., Wilkinson M.~I., 2009, MNRAS, 395, 1146

\bibitem[{Press \& Schechter(1974)}]{Press:1974p2031}
Press W.~H., Schechter P., 1974, ApJ, 187, 425

\bibitem[{Pritchard \& Furlanetto(2007)}]{Pritchard:2007p3787}
Pritchard J.~R., Furlanetto S.~R., 2007, MNRAS, 376, 1680

\bibitem[{Pritchard \& Loeb(2012)}]{Pritchard:2012p2958}
Pritchard J.~R., Loeb A., 2012, Rep. Prog. Phys., 75, 086901

\bibitem[{Ricotti {et~al.}(2001)Ricotti, Gnedin, \& Shull}]{Ricotti:2001p2160}
Ricotti M., Gnedin N.~Y., Shull J.~M., 2001, ApJ, 560, 580

\bibitem[{Ricotti \& Ostriker(2004)}]{Ricotti:2004p7145}
Ricotti M., Ostriker J.~P., 2004, MNRAS, 350, 539

\bibitem[{Santos {et~al.}(2010)Santos, Ferramacho, Silva, Amblard, \&
  Cooray}]{Santos:2010p6501}
Santos M.~G., Ferramacho L., Silva M.~B., Amblard A., Cooray A., 2010, MNRAS,
  406, 2421

\bibitem[{Shaver {et~al.}(1999)Shaver, Windhorst, Madau, \&
  de~Bruyn}]{Shaver:1999p4549}
Shaver P.~A., Windhorst R.~A., Madau P., de~Bruyn A.~G., 1999, A\&A, 345, 380

\bibitem[{Sheth \& Tormen(1999)}]{Sheth:1999p2053}
Sheth R.~K., Tormen G., 1999, MNRAS, 308, 119

\bibitem[{Shimabukuro \& Semelin(2017)}]{Shimabukuro:2017p7140}
Shimabukuro H., Semelin B., 2017, preprint (arXiv:1701.07026)

\bibitem[{Sobacchi \& Mesinger(2014)}]{Sobacchi:2014p1157}
Sobacchi E., Mesinger A., 2014, MNRAS, 440, 1662

\bibitem[{Sokolowski {et~al.}(2015)}]{Sokolowski:2015p6827}
Sokolowski M., {et~al.}, 2015, PASA, 32, e004

\bibitem[{Sun \& Furlanetto(2016)}]{Sun:2016p8225}
Sun G., Furlanetto S.~R., 2016, MNRAS, 460, 417

\bibitem[{Tegmark {et~al.}(1997)Tegmark, Silk, Rees, Blanchard, Abel, \&
  Palla}]{Tegmark:1997p7180}
Tegmark M., Silk J., Rees M.~J., Blanchard A., Abel T., Palla F., 1997, ApJ,
  474, 1

\bibitem[{{Thompson} {et~al.}(2007){Thompson}, {Moran}, \&
  {Swenson}}]{Thompson2007}
{Thompson} A.~R., {Moran} J.~M., {Swenson} G.~W., 2007, {in Interferometry and
  Synthesis in Radio Astronomy. Wiley, New York}

\bibitem[{Thyagarajan {et~al.}(2013)}]{Thyagarajan:2013p2851}
Thyagarajan N., {et~al.}, 2013, ApJ, 776, 6

\bibitem[{Thyagarajan {et~al.}(2015{\natexlab{a}})}]{Thyagarajan:2015p7298}
---, 2015{\natexlab{a}}, ApJL, 807, L28

\bibitem[{Thyagarajan {et~al.}(2015{\natexlab{b}})}]{Thyagarajan:2015p7294}
---, 2015{\natexlab{b}}, ApJ, 804, 14

\bibitem[{Tingay {et~al.}(2013)}]{Tingay:2013p2997}
Tingay S.~J., {et~al.}, 2013, PASA, 30, 7

\bibitem[{Tozzi {et~al.}(2000)Tozzi, Madau, Meiksin, \& Rees}]{Tozzi:2000p4510}
Tozzi P., Madau P., Meiksin A., Rees M.~J., 2000, ApJ, 528, 597

\bibitem[{Trott {et~al.}(2012)Trott, Wayth, \& Tingay}]{Trott:2012p2834}
Trott C.~M., Wayth R.~B., Tingay S.~J., 2012, ApJ, 757, 101

\bibitem[{Tzanavaris \& Georgantopoulos(2008)}]{Tzanavaris:2008p8147}
Tzanavaris P., Georgantopoulos I., 2008, A\&A, 480, 663

\bibitem[{van Haarlem {et~al.}(2013)}]{vanHaarlem:2013p200}
van Haarlem M.~P., {et~al.}, 2013, A\&A, 556, 2

\bibitem[{Vedantham {et~al.}(2012)Vedantham, Shankar, \&
  Subrahmanyan}]{Vedantham:2012p2801}
Vedantham H., Shankar N.~U., Subrahmanyan R., 2012, ApJ, 745, 176

\bibitem[{Voytek {et~al.}(2014)Voytek, Natarajan, Garc{\'\i}a, Peterson, \&
  L{\'o}pez-Cruz}]{Voytek:2014p6741}
Voytek T.~C., Natarajan A., Garc{\'\i}a J. M.~J., Peterson J.~B.,
  L{\'o}pez-Cruz O., 2014, ApJL, 782, L9

\bibitem[{Watkinson \& Pritchard(2015)}]{Watkinson2015a}
Watkinson C.~A., Pritchard J.~R., 2015, MNRAS, 454, 1416

\bibitem[{Wouthuysen(1952)}]{Wouthuysen:1952p4321}
Wouthuysen S.~A., 1952, AJ, 57, 31

\bibitem[{Xu {et~al.}(2014)Xu, Ahn, Wise, Norman, \& O'Shea}]{Xu:2014p6671}
Xu H., Ahn K., Wise J.~H., Norman M.~L., O'Shea B.~W., 2014, ApJ, 791, 110

\bibitem[{Xu {et~al.}(2016)Xu, Wise, Norman, Ahn, \& O'Shea}]{Xu:2016p1}
Xu H., Wise J.~H., Norman M.~L., Ahn K., O'Shea B.~W., 2016, ApJ, 833, 84

\bibitem[{Yatawatta {et~al.}(2013)}]{Yatawatta:2013p2980}
Yatawatta S., {et~al.}, 2013, A\&A, 550, A136

\bibitem[{Yue {et~al.}(2013)Yue, Ferrara, Salvaterra, Xu, \&
  Chen}]{Yue:2013p6705}
Yue B., Ferrara A., Salvaterra R., Xu Y., Chen X., 2013, MNRAS, 433, 1556

\bibitem[{Zahn {et~al.}(2011)Zahn, Mesinger, McQuinn, Trac, Cen, \&
  Hernquist}]{Zahn:2011p1171}
Zahn O., Mesinger A., McQuinn M., Trac H., Cen R., Hernquist L.~E., 2011,
  MNRAS, 414, 727

\bibitem[{Zaroubi(2013)}]{Zaroubi:2013p2976}
Zaroubi S., 2013, {The First Galaxies, Astrophysics and Space Science Library,
  Vol. 396. Springer-Verlag, Berlin, p. 45}

\bibitem[{Zel'dovich(1970)}]{Zeldovich:1970p2023}
Zel'dovich Y.~B., 1970, A\&A, 5, 84

\end{thebibliography}

\appendix

\section[]{Sensitivity Curves} \label{sec:PSfigures}

\begin{figure*} 
	\begin{center}
		\includegraphics[trim = 0cm 0.8cm 0cm 0.6cm, scale = 0.72]{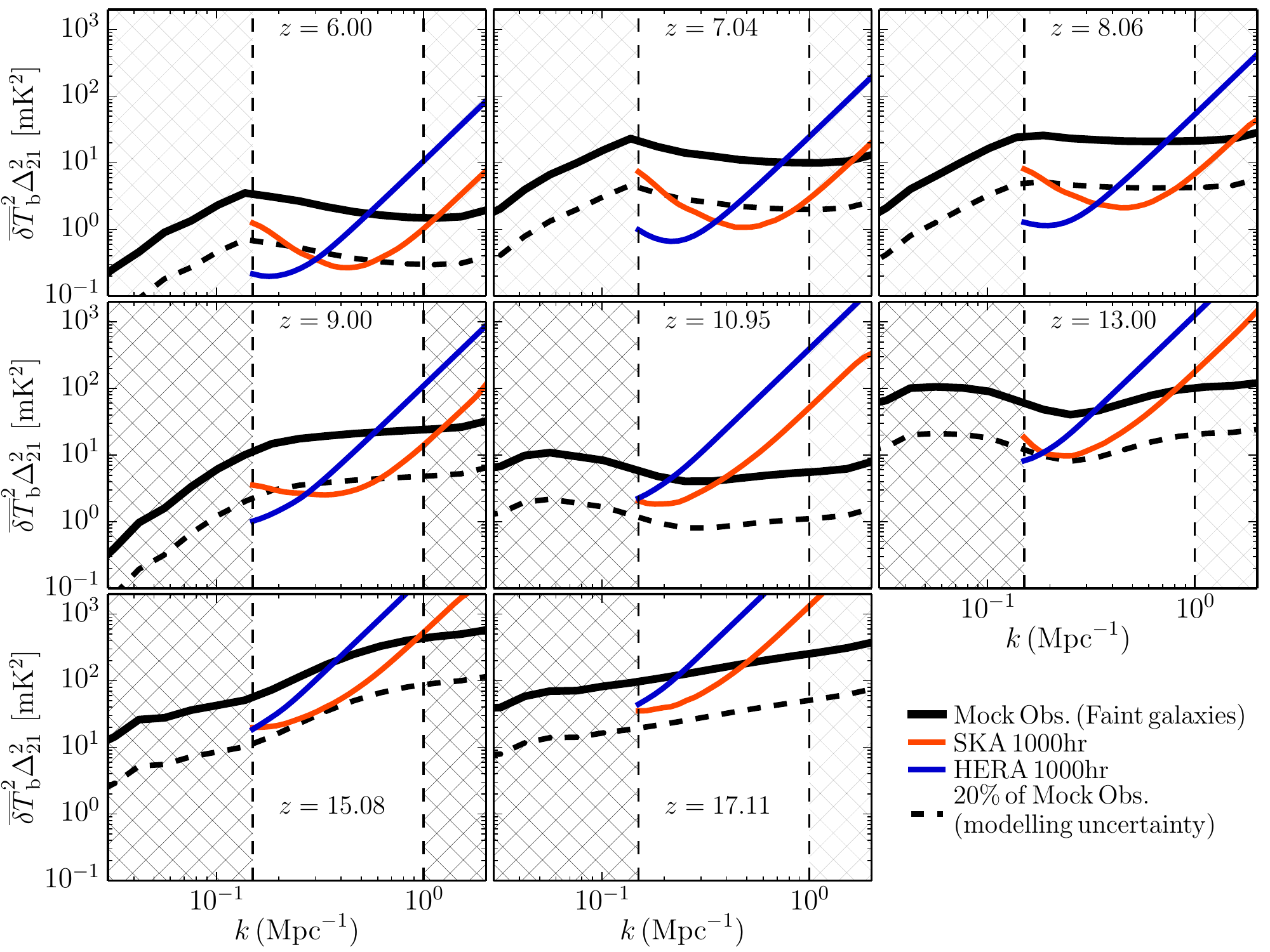}
	\end{center}
\caption[]
{The 21 cm PS (black, solid curve) of the \textsc{faint galaxies} (Section~\ref{sec:MockObs}) mock observation at all eight co-eval redshifts used in this work. Red and blue solid curves correspond to the noise curves (thermal + sample variance) for an assumed 1000 hr observation with the SKA and HERA, respectively (see Section~\ref{sec:noise}). The black dashed curve is 20 per cent of the mock 21 cm PS, highlighting the scale of the assumed modelling uncertainty used in this work. Representing it in this format highlights the dominant source of error at any redshift, or $k$-mode. Hatched regions correspond to $k$-modes beyond the 21 cm PS fitting region used in \cmmc{}.}
\label{fig:FaintGalaxies_Sensitivity}
\end{figure*}

\begin{figure*} 
	\begin{center}
		\includegraphics[trim = 0cm 0.8cm 0cm 0.6cm, scale = 0.72]{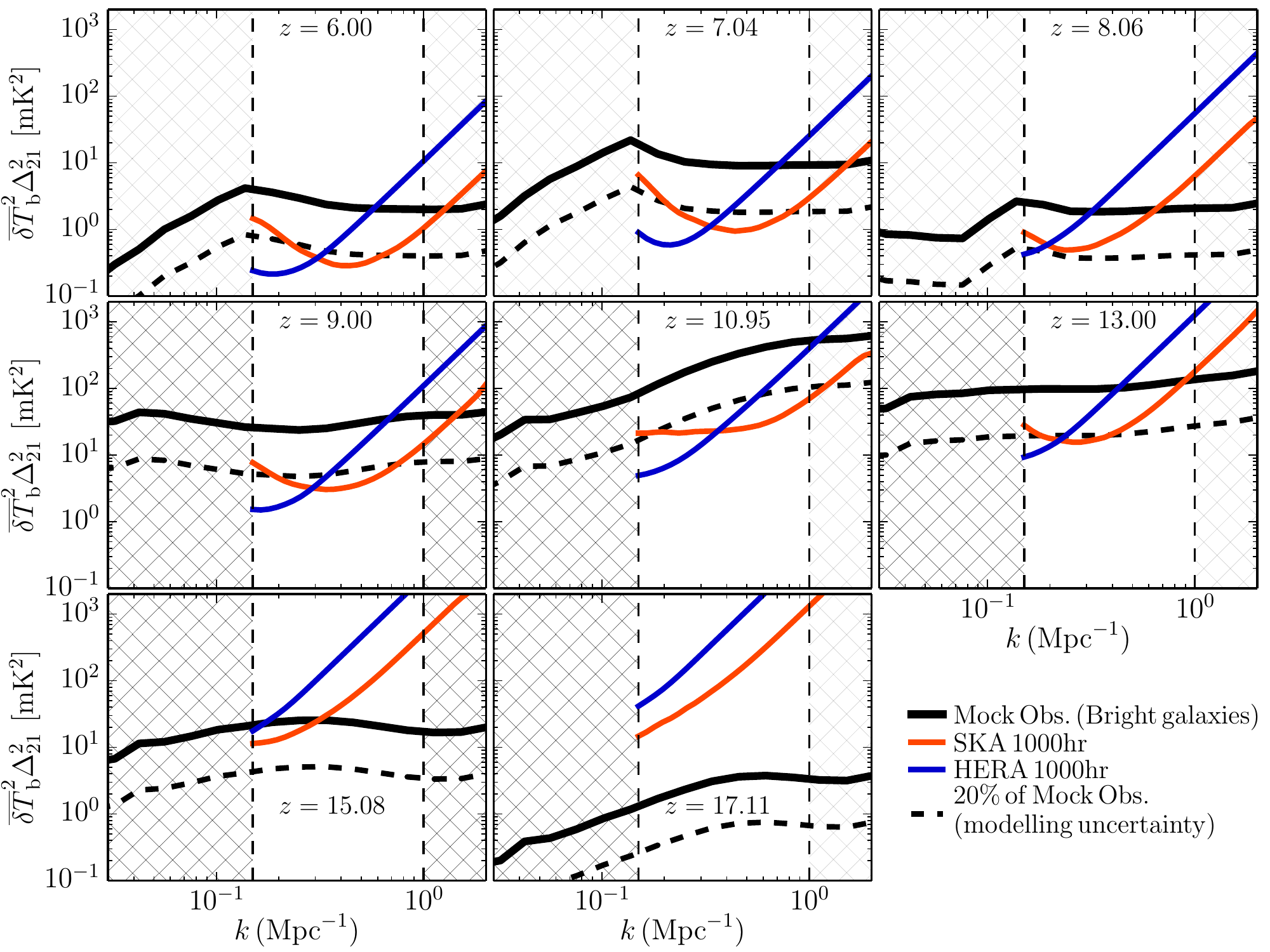}
	\end{center}
\caption[]
{Same as Figure~\ref{fig:FaintGalaxies_Sensitivity} except now for the \textsc{bright galaxies} model (see Section~\ref{sec:MockObs}).}
\label{fig:BrightGalaxies_Sensitivity}
\end{figure*}

In Figures~\ref{fig:FaintGalaxies_Sensitivity} and~\ref{fig:BrightGalaxies_Sensitivity} we present the mock 21 cm PS (black curves) for both of our EoR source models \textsc{faint galaxies} and \textsc{bright galaxies}, respectively. We show the 21 cm PS for each of the eight co-evolution redshifts we use in this work to achieve our astrophysical constraints, covering both the EoR and the EoH. For each, we provide the corresponding total noise PS for HERA (blue) and SKA (red) for our assumed 1000hr observational strategy (see Section~\ref{sec:noise} for more details). Finally, the black dashed curve is the 20 per cent modelling uncertainty, which we have applied to the mock 21 cm PS (it is applied to the sampled 21 cm PS not the mock observation in \cmmc{}) to provide reference to the dominate source of error at any $k$-mode.

It is interesting to note that while most of the EoH parameters are tightly constrained for both mock observations,
these constraints appear to be arise solely from a relatively narrow range in redshifts, which pick up the rise and fall of the large scale power. This emphasises the importance of observing the cosmic 21 cm signal in an extended frequency range, allowing us to pick up the major milestones in the signal which drive strong astrophysical parameter constraints.

\end{document}